\documentclass[prd,twocolumn,amsmath,amssymb,showpacs,floatfix,nofootinbib,preprintnumbers,superscriptaddress,aps,longbibliography]{revtex4-2}

\usepackage[latin1]{inputenc}
\usepackage[normalem]{ulem}
\usepackage[mathscr]{euscript}
\usepackage{todonotes}
\usepackage{amsmath}
\usepackage{printlen}
\usepackage{amsfonts}
\usepackage{amssymb}
\usepackage{dcolumn}
\usepackage{graphicx}
\usepackage{tikz}
\usetikzlibrary{shapes.geometric, arrows}
\tikzstyle{arrow} = [ultra thick,->,>=stealth]
\usepackage{orcidlink}
\usepackage{wrapfig}
\usepackage{siunitx}
\usepackage{appendix}
\usepackage{bm}
\usepackage{ulem}
\usepackage{latexsym}
\usepackage{epsfig}
\usepackage{hyperref}
\hypersetup{colorlinks=true}
\usepackage{soul,xcolor}
\setstcolor{red}
\usepackage{ragged2e}
\usepackage[caption=false]{subfig}

\begin{document}
\title{SCoRe: A New Framework to Study Unmodeled Physics \\ from Gravitational Wave Data}

\date{\today}

\author{Guillaume Dideron \orcidlink{0000-0002-5222-7974}}\email{gdideron@perimeterinstitute.ca}
\affiliation{Department of Physics and Astronomy, University of 
Waterloo, 200 University Avenue West, Waterloo, ON N2L 3G1, Canada}
\affiliation{Perimeter Institute for Theoretical Physics, 31 Caroline Street North, Waterloo, ON N2L 2Y5, Canada}

\author{Suvodip Mukherjee \orcidlink{0000-0002-3373-5236}}\email{suvodip@tifr.res.in}
\affiliation{Department of Astronomy \& Astrophysics, Tata Institute of Fundamental Research, 1, Homi Bhabha Road, Colaba, Mumbai 400005, India}

\author{Luis Lehner \orcidlink{0000-0001-9682-3383}}\email{llehner@perimeterinstitute.ca}
\affiliation{Perimeter Institute for Theoretical Physics, 31 Caroline Street North, Waterloo, ON N2L 2Y5, Canada}

\begin{abstract}
	A confident discovery of physics beyond what has been consistently
modeled from gravitational wave (GW) data requires a technique that can
distinguish between noise artifacts and unmodeled signatures while also
shedding light on the underlying physics. We propose a new data analysis
method, \texttt{SCoRe} (Structured Correlated Residual), to search for
unmodeled physics in GW data, which addresses both of these aspects. The method
searches for structure in the cross-correlation power spectrum of residual
strains of pairs of GW detectors by projecting this power spectrum onto a
frequency-dependent template. The template may be model-independent or
model-dependent and is constructed based on the properties of the GW source
parameters. The projection of the residual strain enables distinction between
noise artifacts and any true signal while capturing possible dependence on the
GW source parameters. Our method is constructed within a Bayesian framework,
and we demonstrate its application on a model-independent toy example and for a
model motivated by an effective field theory of gravity. The method developed
here will be useful for searching for a wide variety of new physics and
yet-to-be-modeled known physics in GW data accessible from the current network
of LIGO-Virgo-KAGRA detectors, as well as from future earth- and space-based GW
detectors such as A+, LISA, Cosmic Explorer, and the Einstein Telescope.
\end{abstract}

\maketitle

\section{Introduction}\label{intro}

The direct detection of gravitational waves (GWs) by the LIGO-Virgo
collaboration~\cite{ligoscientificcollaborationandvirgocollaborationObservationGravitationalWaves2016a}
and the discovery of nearly 100 GW sources since
then~\cite{theligoscientificcollaborationGWTC1GravitationalWaveTransient2019,abbottGWTC2CompactBinary2021,theligoscientificcollaborationGWTC3CompactBinary2021}
has opened a new avenue to study fundamental physics using compact binaries in
relativistic regimes. Binary Neutron Stars (BNSs), Neutron Star-Black Holes
(NSBHs), and Binary Black Holes (BBHs) provide unprecedented opportunities to
explore the nature of gravity over a vast range of mass scales and cosmological
distances. Among the many intriguing questions that can be addressed with GWs
is whether data from the LIGO-Virgo-KAGRA (LVK) and future detectors can reveal
deviations from our current Standard Model (SM) for GWs. This model includes
(non-exhaustively) our best theory of gravity - General Relativity (GR), black
holes and/or neutron stars describing compact objects, and the extent to which
relevant physical features can be consistently accounted for in our waveform
templates (for example, how well certain ranges of parameters, such as high
eccentricity or high mass ratios, can be captured by numerical template banks),
along with our understanding of the noise in the GW detection facilities. We
will refer to any deviation from this combination of assumed knowledge as a
Beyond Modeled (BM) signature. For each event, the standard model is used to
produce a best-fit waveform and, crucially, to interpret the physical content
of the signal. Current tests for BM signatures examine potential deviations
from this best-fit waveform.

Null tests are arguably the cleanest approach to testing the current SM.
They determine whether signals are consistent with the expectations of the 
model. However, on their own, they do not inform us of the physical meaning of
a given deviation. To that end, other tests have been devised that target
parameterized deviations in the Post-Newtonian (and Post-Einsteinian)
description of the inspiraling behavior and/or quasi-normal modes of the newly
formed black hole that might result from a merger (see
e.g.~\cite{theligoscientificcollaborationTestsGeneralRelativity2021a,agathosTIGERDataAnalysis2014a}
and references cited therein). With these tests, such deviations can, in
principle, be closely connected with specific effects from particular theories
- or at least provide physical information on the main impact of deviations
from GR~\cite{Yunes:2009ke}. On the other hand, agnostic tests examine the
residual (between the consistent SM signal and the observed data) and aim to
qualify the behavior of any departure (see
e.g~\cite{theligoscientificcollaborationTestsGeneralRelativity2021a,Johnson-McDaniel:2021yge}).\footnote{Rather
	than looking at the difference between the data and the best-fit
	waveform template, agnostic tests can also add deviations with general
forms directly to the waveform templates, such as in~\cite{Edelman_2021}, where
splines are used to fit the deviations.} This approach can accommodate
yet-to-be-modeled physics. At this point, we find it worthwhile to highlight an
obvious fact: eventually, for sufficiently high SNR, all models (even within
the SM) will show limitations due to systematics, such as physics still to be
computed, yet unknown physical ingredients to be accounted for, or intrinsic
waveform modeling errors. This is particularly true for future GW detectors.
Their sensitivity will allow thousands of events to be combined to test for the
presence of a BM signal. Even small systematic errors in waveform models can
then accumulate and cause the misidentification of a BM deviation as shown
in~\cite{huAccumulatingErrorsTests2022}. The problem is amplified when events
overlap, which may frequently occur in future detectors. Thus, devising
suitable strategies to analyze residuals will become increasingly important.

To date, available observations obtained with the LVK network have already made
it possible to put constraints on BM phenomenology. For instance, by
constraining beyond-GR gravity and potential effects associated with exotic
compact objects. For the former, some constraints have been derived on
deviations with particular physical consequences. For example, constraints have
been derived on the coupling parameters of certain Scalar-Tensor theories in
neutron star
binaries~\cite{lyuConstraintsEinsteindilationGaussBonnetGravity2022,niuConstrainingScalartensorTheories2021},
the graviton
mass~\cite{theligoscientificcollaborationTestsGeneralRelativity2021a}, and the
number of possible space-time dimensions~\cite{pardoLimitsNumberSpacetime2018}.
Others have been derived on deviation coefficients that may arise in many
different beyond GR theories; for example on the post-Newtonian expansion
coefficients~\cite{nairFundamentalPhysicsImplications2019,
yunesTheoreticalPhysicsImplications2016}, on propagation
effects~\cite{kosteleckyTestingLocalLorentz2016,Mastrogiovanni:2020gua,
okounkovaConstrainingGravitationalWave2022, shaoCombinedSearchAnisotropic2020},
on additional modes of polarizations, and on some potential signatures of black
hole mimickers~\cite{ abediEchoesAbyssTentative2017,
	ashtonCommentsEchoesAbyss2016, uchikataSearchingBlackHole2019,
	westerweckLowSignificanceEvidence2018}.\footnote{ Frameworks for
		obtaining these constraints are developed
		in~\cite{willBoundingMassGraviton1998} for the mass of the
		graviton,
		in~\cite{belgacemGravitationalwaveLuminosityDistance2018,Belgacem:2018lbp,nishizawaGeneralizedFrameworkTesting2018,Mukherjee:2020mha}
		and
		\cite{mewesSignalsLorentzViolation2019,mirshekariConstrainingLorentzviolatingModified2012,oneal-aultAnalysisBirefringenceDispersion2021}
		for propagation effects, and for exotic compact objects by
		analysing the post-merger
		ringdown~\cite{bertiGravitationalwaveSpectroscopyMassive2006,
			cardosoGravitationalWaveRingdownProbe2016,cardosoGravitationalwaveSignaturesExotic2016,
			cardosoTestsExistenceBlack2017,gossanBayesianModelSelection2012,meidamTestingNohairTheorem2014,
		dreyerBlackholeSpectroscopyTesting2004}. Some descriptions of
		possible exotic compact objects
		include~\cite{mazurGravitationalVacuumCondensate2004,
		lieblingDynamicalBosonStars2012,
	giudiceHuntingDarkParticles2016,
danielssonDynamicsObservationalSignatures2021}. The use of detectors to search
for non-GR polarization modes was described as far back as
1973~\cite{eardleyGravitationalWaveObservationsTool1973}.} Constraints for all
of the effects mentioned above have been presented {by the LVK
collaboration}~\cite{ligoscientificcollaborationandvirgocollaborationTestsGeneralRelativity2019,theligoscientificcollaborationTestsGeneralRelativity2021a}.

Notably, no compelling evidence has so far been found for deviations from GR.
However, 
as the sensitivity of GW detectors
improves~\cite{Yang:2017zxs,Zimmerman:2019wzo,Kalogera:2021bya}, new
ground- \cite{Kalogera:2021bya}
and space-based~\cite{Barausse:2020rsu} detectors become available,
and more events are detected, increasingly deeper tests will be possible. One
key challenge in identifying a BM signature from the GW data is 
instrumental noise, especially in light of current evidence suggesting any
deviation is likely to be subtle. Indeed, even in the scenario where the
statistical properties of a detector are well characterized, it is difficult to
confidently assign the origin of a small BM signature as astrophysical and not
from a known (or unknown) instrumental property. This is because the noise
property of a detector can be far from a Gaussian distribution and it can also
have non-stationary
behavior~\cite{abbottCharacterizationTransientNoise2016,abbottGuideLIGOVirgo2020,blackburnLSCGlitchGroup2008,chatziioannouNoiseSpectralEstimation2019,richabbottOpenDataFirst2021,edyIssuesMismodelingGravitationalwave2021}.
In particular, glitches and other uncontrolled sources of noise make searching
for BM signatures challenging~\cite{kwokInvestigationEffectsNonGaussian2022}.
Distinguishing the BM signature from detector noise is only the
first challenge---as then one would also like to understand the physics of such
signature and its dependence on the GW source properties.

Motivated by the above considerations, we propose a new method
called \textit{Structured Correlated Residual} (\texttt{SCoRe}) power spectrum.
\texttt{SCoRe} searches for BM signatures by measuring the Correlated Residual
Power Spectrum (CRPS) between multiple pairs of GW detectors. This method is
designed such that it is not susceptible to contribution from uncorrelated
noise between a pair of GW detectors and can search for any (un-)modeled
signals.  

\texttt{SCoRe} applies cross-correlation, a technique used in searches for
stochastic GW background~\cite{ligo_scientific_collaboration_analysis_2004,
allen_detecting_1999} to the residual data obtained after subtracting a
\textit{best-fit} model of the signal from the strain data. This best-fit model
is constructed using the values allowed by the posterior on the GW source
parameters inferred from the detector network using the SM.
The method then searches for a specific type of frequency dependence
in the CRPS which can be unmodeled and driven by \textit{chirp-like} behavior,
or model-specific. The cross-correlation technique ensures that
uncorrelated sources of noise will not contribute to the mean of the CRPS.
Meanwhile, projection to a suitable chosen set of functions can help pick
features missing from the SM for follow-up analysis on their physical meaning.
This is described in detail in Sec.~\ref{formalism}. It is important to note
that any source of correlated noise, such as the noise that arises due to the
Schumann
resonance~\cite{schumannUberDampfungElektromagnetischen1952,nguyenEnvironmentalNoiseAdvanced2021,thraneCorrelatedMagneticNoise2013,
thraneCorrelatedNoiseNetworks2014}, will contaminate the cross-correlation
signal. We do not consider correlated noise in this work. Future studies will
investigate the impact of correlated noise on \texttt{SCoRe} and
how it can be mitigated.

In this work, we consider the ``best-fit" model of the signal to be the maximum
likelihood GR waveform obtained by parameter estimation of the network strain.
In the presence of a BM signature, this estimator may have a ``stealth bias".
This is when a GR waveform with different source parameters can --even
partially-- account for the
signature~\cite{vallisneriStealthBiasGravitationalwave2013,vitaleHowSeriousCan2014}.
In the worst scenario, a different set of GR parameters may be able to
perfectly reproduce the BM signature. It would then be impossible for any
residual test to distinguish such BM signatures from the SM.
In~\cite{okounkovaGravitationalWaveInference2022}, the effect of stealth biases
on the measurement of a particular BM model has been studied. The model
consisted in Numerical Relativity (NR) waveforms produced through an order
reduced strategy designed to capture, to a certain extent, dynamical
Chern-Simons gravity with source parameters similar to a detected signal,
GW150914~\cite{ligoscientificcollaborationandvirgocollaborationObservationGravitationalWaves2016a}.
Although the residual strain could partially be reproduced by GR waveforms, it
was found that at least part of it could not be accounted for by the GR source
parameters correcting for the BM signature, a consequence of deviations lying
off the GR waveforms manifold. We note in passing that it is not guaranteed
that all BM models have an orthogonal component to the SM manifold.

In fact, there exist beyond GR theories allowing for signals purely degenerate
with GR waveforms for some source parameters and not for others. For example,
theories that predict larger deviations at higher curvature scales may not lead
to measurable differences with GR for larger compact object masses. Another
well-known example is spin-induced
moment~\cite{ryanGravitationalWavesInspiral1995,poissonGravitationalWavesInspiraling1998,laarakkersQuadrupoleMomentsRotating1999,krishnenduTestingBinaryBlack2017},
which by definition is an effect proportional to the spin of the object.
We do not explore such degenerate scenarios in this work.

This work is organized as follows. In Sec. \ref{tgr}, we describe the
motivation behind this new technique. In Sec.~\ref{formalism} and
Sec.~\ref{results_1}, we discuss the formalism of the work and its application
to a simple toy example. In Sec.~\ref{results_2}, we show the application of
this method to a specific theoretical model. Finally, we discuss the
conclusions and future outlooks in Sec.~\ref{conc}.  

In this work, we develop the mathematical framework of this new technique
\texttt{SCoRe} and tested its performance for Gaussian stationary noise. In
future work, we will consider its application on a population of GW sources
detected by the LVK collaboration and will include non-stationary and
non-Gaussian noise.

\begin{figure*}[!ht]
  \includegraphics{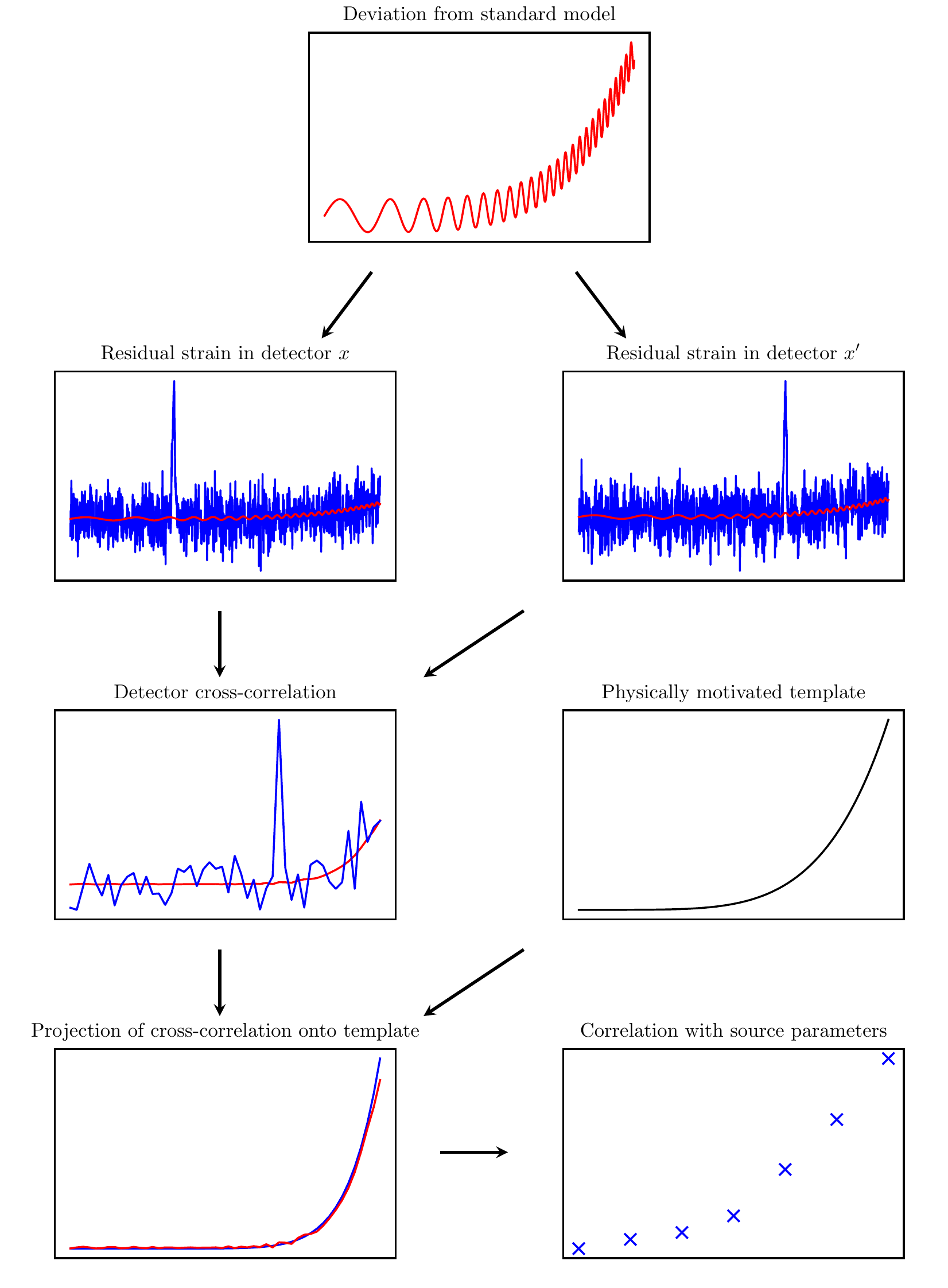}
  \caption{Outline of the \texttt{SCoRe} method. When a BM signature (in red) is
  measured by a pair of GW detectors, each detector will add onto it a
  different noise. The resulting residual strains, computed by subtracting the
  best-fit SM waveform from the data, are plotted in blue. Noise that is
  uncorrelated between detectors will be reduced when taking the
  cross-correlation between the two detectors. Higher power transient noise and
  correlated noise may remain, but can be mitigated by projecting 
  CRPS onto some physically motivated template. Most BM signature models
  have a dependence on the source parameters. The projection
  coefficients measured over events of different source parameters may reveal
	such correlation. }
  \label{fig:method_flowchart}
\end{figure*}

\section{Motivation behind {SCoRe}}\label{tgr}
The search for BM physics using GWs is challenging due
to multiple reasons, including both theoretical and experimental limitations.
On one side, we are limited by the number of theoretical waveforms needed to
test for the vast range of potentially BM scenarios (either for missing
physics within GR, or sufficiently generic and complete waveforms beyond it). On the other side, we are limited by
our understanding of the instrument noise properties when differentiating any
BM signature from noise artifacts. Thus, often one fits the data with available waveform
models to infer the best-fit source
parameters and searches for any coherent residual
signal~\cite{abbottPropertiesAstrophysicalImplications2020,ligoscientificandvirgocollaborationsTestsGeneralRelativity2016,theligoscientificcollaborationandthevirgocollaborationtestsgeneralrelativity2019a,theligoscientificcollaborationTestsGeneralRelativity2021a,theligoscientificcollaborationTestsGeneralRelativity2021}.
However, the limitation remains in how well a framework can identify a BM signature and
what the appropriate metric to quantify a deviation is. An example
of this is the uneven performance of specific tests with different
events (e.g.~\cite{Johnson-McDaniel:2021yge}).

A few salient features expected from any BM signature(s) in the GW data are (i)
BM signatures will be {common in the data of multiple detectors, but
uncorrelated detector noise will not}, (ii) a large class (if not all) of BM
signatures can be understood as additional loss/gain in the orbital energy of a
coalescing binary (e.g.~\cite{Damour:1992we,Yunes:2009ke,Barausse:2012da,Will:2014kxa}) when 
compared to the SM (GR, and sources involving black holes
and/or neutron
stars) , (iii) BM
signatures can depend on the GW source properties.

In this work, we propose a new technique to search for BM signatures in GW data
that aims to address the issues mentioned above. This method has three key
characteristics, each exploiting and/or targeting the three of BM signatures
listed previously: (i) the residual GW data obtained after subtracting
a modeled GW signal for a given waveform and source parameters are
cross-correlated within pairs of detectors --to help distinguish 
signal from noise; (ii) the CRPS is projected on a physically motivated
template. This template is either composed of the derivatives of the orbital
frequency up to order $n$ (for a model-independent search), or derived from a
particular physical model (for a model-dependent search) --to help characterize
the signal; (iii) the projection is estimated as a function of the GW best-fit
source parameters from the signal after marginalizing its uncertainties. 

The first step helps in mitigating the contribution from uncorrelated noise
to the residual signal by cross-correlating a total of $N_{\rm
det}(N_{\rm det}-1)/2$ pairs of detectors, where $N_{\rm det}$ is the
number of detectors.
Cross-correlation is a useful tool technique to search for weak signals from
noisy GW data and is often used to search for stochastic GW background signals
~\cite{ligo_scientific_collaboration_analysis_2004, allen_detecting_1999}.  It
is a time averaging over the product of the strains in two different detectors.
Because the noise at the two detectors is not correlated, its mean average is
zero, and its standard deviation decreases as the timescale of integration is
increased. A price is paid however for this noise reduction: features shorter
than the averaging timescale will be lost. 
The second step aids to look for a
\textit{physics-driven} BM signal in either a model-dependent
or a model-independent way. 
Projecting onto template functions also
further reduces noise by filtering particular, model-motivated, structures in
the signal. We suggest templates that are naturally expressible in terms of the
orbital frequency. The final step takes into account possible
dependencies of the BM signal on the GW source parameters and can also
mitigate uncertainties related to the source parameters. 

Although the idea of using template functions in frequency space is similar to
the frames used in the \texttt{BayesWave}
framework~\cite{cornishBayeswaveBayesianInference2015,cornishBayesWaveAnalysisPipeline2021}
(which can also be used as template functions in \texttt{SCoRe}), our method
does not seek to reproduce the residual signal with a set of functions.  This
is unlike current residual tests that use
\texttt{BayesWave}~\cite{theligoscientificcollaborationTestsGeneralRelativity2021,theligoscientificcollaborationTestsGeneralRelativity2021a}.
Instead, we measure the degree to which a pre-determined finite set of
functions can explain the residual strain. This makes \texttt{SCoRE}
morphology-dependent, but the morphology of the chosen template functions is
arbitrary and can be non-specific. For example, one could reproduce the
deviations parameters  used in parametrized
tests~\cite{agathosTIGERDataAnalysis2014a,PhysRevD.97.044033,mehtaTestsGeneralRelativity2022,theligoscientificcollaborationTestsGeneralRelativity2021},
but choosing an agnostic template set using \texttt{SCoRe} is also possible, as
we will explain in Sec~\ref{sec:formalism_basis}.

By using cross-correlation and projection onto a template, we hope to reduce
noise well below that of individual detectors. With this noise reduction,
\texttt{SCoRe} may be able to unearth minute BM signatures, which would
otherwise remain undetected by only considering the strain of individual
detectors.  
By looking at the structure of the CRPS, the method can help constrain potential deviations with a vast range of possible forms. This also
means that the search for potential deviations can be made tighter
as their form is guided with theoretical
work.
Furthermore, the method can also probe for
correlations between the BM signature and the GW source parameters. A schematic
diagram illustrating the basic principles of this technique is depicted in
Fig.~\ref{fig:method_flowchart}.

\section{Formalism for SCoRe}\label{formalism} We describe below the formalism
of \texttt{SCoRe}, which consists of four parts, \textbf{(A) Cross-correlation,
(B) Choice of a {residual} template, (C) Projection on a template, and  (D)
Inference using a Bayesian framework}.
We can write a linear model of the GW data $ d^x(t)$ in a detector denoted by
$x$ as 
\begin{align}
\begin{split}
    d^x(t)=& s^x(t) + n^x(t),\\
\end{split}
\end{align}
where $s^x(t)= \sum_i F^x_i(t)h_i(t)$ denotes the observed GW signal as seen by
a detector, written in terms of the actual signal $h_{i}(t)$ for the
polarization state ``$i$", and the corresponding detector response function
$F^x_i$(t). The noise in detector $x$ is $n^x(t)$. 
Whether a signal is detected
in a particular detector depends on the noise in the detector and the detector response function
$F^x_i$(t) to the source position.
In this work, we consider that
a GW signal detected with a SM template in
multiple detectors with a matched filtering SNR $\rho$ greater than a minimum
threshold network SNR $\rho_*$ is classified as an event.\footnote{See~\cite{theligoscientificcollaborationGWTC1GravitationalWaveTransient2019,abbottGWTC2CompactBinary2021,theligoscientificcollaborationGWTC3CompactBinary2021} for a description of the actual pipelines used by the LVK to identify candidates.}
Performing parameter estimation over the strains in the
detector network gives a posterior over the source parameters $\theta$. By
definition, the Maximum Likelihood Estimate (MLE), $\theta_\text{MLE}$,
corresponds to the minimum residual strain amplitude
in comparison to the variance in the detector noise, as $\theta_\text{MLE}$
should maximize the likelihood. 

We propose to perform the test of BM physics on the residual data which is
constructed using the MLE parameters inferred using a network of detectors,
$\theta_\text{MLE}$ as
\begin{align}
\begin{split}
    \Delta d_{\theta_\text{MLE}}^x(t)=& s(t)  - s_{BF}(t) + n^x(t), \\
     =& \Delta s(t) + n^x(t),
     \end{split}
\end{align}
where $s_{BF}(t)$ is the SM waveform for $\theta_\text{MLE}$. 
The MLE parameters need not be the
true parameters of the source parameters---they only resemble the data best
given a waveform model and the assumption of the standard model.
{Since such parameters will be generically described with a distribution, we introduce in Sec.~\ref{Bayesian_framework} a Bayesian framework to 
marginalize their (posterior) distribution.}

\subsection{Cross-correlation of the residual signal}

One of the key aspects of a BM signature in the residual data is that it is likely to be
correlated in all the detectors. However, it will be observed differently in each detector depending on the detector response function.
Noise between multiple GW detectors
from unrelated sources will be random. To identify the origin of the residual
as either astrophysical (a BM signature) or due to detector noise, we propose
to do a cross-correlation of the residual signal between the data from two
detectors. Given the output $d^{x}(t)$ and $d^{x'}(t)$ of two detectors $x$ and
$x'$, the general cross-correlation $Y(\tau)$ of the two signals over a
timescale $\tau$ is defined as 
{
\begin{align}
   \label{eq:cc}
   &
	    Y(\tau)
     \nonumber\\
    &\equiv
    \frac{1}{\tau}
    \int_{t-\tau/2}^{t+\tau/2} dt_1
    \int_{t-\tau/2}^{t+\tau/2} dt_2
    d^{x}(t_1) d^{x'}(t_2)Q(t_1,t_2),
\end{align}
}
where $Q(t_1,t_2)$ is a filter function.
The choice of $\tau$ should balance two effects: correlated noise is reduced
better as $\tau$ increases, but averaging washes away signal features with
timescale shorter than $\tau$.
{
In the limit where $\tau$ spans the length of
the signal, the cross-correlation becomes the total residual power observed.
}
There are natural choices of $\tau$, which rely on maximizing SNR while
ensuring the features of the signal are preserved. They will depend on the
source parameters, most importantly the chirp mass, which sets the timescale of
the orbital frequency evolution, and the choice of template function on which
to project the CRPS. We discuss the methodology behind choosing $\tau$ with a
specific example in Sec.~\ref{sec:choice_of_tau}.

The definition~\eqref{eq:cc} has been used to search for stochastic GW
signal~\cite{ligo_scientific_collaboration_analysis_2004,allen_detecting_1999}.
There, the strain in multiple detectors is used to search for a common,
periodic signal ~\cite{Dhurandhar:2007vb}. In stochastic GW background
searches, the phase of the signal is
not expected to be resolved. This makes choosing the optimal filter function
non-trivial. By contrast,  we can measure the phase of a well-detected event and
measure $\Delta t$, the delay time separating the two detectors {within a timing uncertainty}.
The optimal filter function, which gives the highest SNR, will correspond to
the signals in the detectors overlapping in time.  It is therefore $Q(t_1,t_2)=\delta(|t_1-t_2|-\Delta
t)$ (assuming $\tau \gg \Delta t$).  
We can then define the cross-correlation with maximum SNR with angular brackets:
\begin{align}
	\label{eq:cc_definition}
	\begin{split}
	&\left\langle d^{x}(t) d^{x'}(t + \Delta t) \right\rangle 	\\
	&\equiv
	\frac{1}{\tau}	
	\int_{t-\tau/2}^{t+\tau/2} dt'
	d^{x}(t') d^{x'}(t' + \Delta t),
	\end{split}
\end{align}
and, more specifically, the cross-correlation of the residual strains:
\begin{align}
	\begin{split}
		D^{xx'}(t)
		\equiv &
		\langle 
		\Delta d_{\theta_\text{MLE}}^x(t) \Delta d_{\theta_\text{MLE}}^{x'}(t+\Delta t) 
		\rangle \label{eq:D_definition}, \\
		=&
		\langle 
		\Delta s^x_{\theta_\text{MLE}}(t) \Delta s^{x'}_{\theta_\text{MLE}}(t+\Delta t) 
		\rangle 
		+
		\langle
		n^{x}(t) n^{x'}(t+\Delta t)
		\rangle.
	\end{split}
\end{align}
The value of $\Delta t$ will depend on the source parameters such as the sky
position and also on the position of the two GW detectors $x$ and $x'$ between
which the signal is correlated. This can be inferred for individual pairs of
detectors, and need not be assumed in the analysis. The detection of the correlated
signal $D(t)$ depends on the overlap of the signal $\Delta s^x_{\theta_{\rm
MLE}}(t)$ between the individual detectors, which depends on the 
response functions of the detectors $F_i^x$(t). 
We can marginalize over the response functions and extrinsic source parameters, as
explained in Sec.~\ref{Bayesian_framework}, so that we may account for
different residual signatures that will be induced by the same BM deviation in
different detectors. But, if, for a given antenna pattern, a theory predicts no
measurable residual signature in multiple detectors, then the cross-correlation
will not yield anything.

Using the cross-correlation definition given in Eq.~\eqref{eq:cc_definition},
we can model
the measured cross-correlated signal as 
\begin{equation}\label{eqcc1}
   D^{xx'}(t)
   =
   S^{xx'}_{\theta_\text{MLE}}(t)
   +
   \mathcal{N}^{xx'}_c(t),
\end{equation}
where the first term denotes the correlated signal between the two detectors
and the second term denotes the correlated residual noise between the two
detectors. We usually assume that the noise between detectors is uncorrelated. 
In reality, some noise, like magnetic noise, can be correlated. Though these types
of noise are unlikely to be dominant, they will play a role at some level in the mean, and their impact on the signal should be estimated when this technique is applied to the data. 

If the noise is uncorrelated between the two detectors, then the second term in
Eq.~\eqref{eqcc1}
vanishes and only the first term is non-zero. As a result, we can
write 
\begin{align}\label{cc1}
    D^{xx'}(t)
    =&
    \langle
    	\Delta s^{x}_{\theta_\text{MLE}}(t) \Delta s^{x'}_{\theta_\text{MLE}}(t+\Delta t) 
    \rangle \nonumber
    \\ &+
    \langle n^{x}(t) n^{x'}(t+\Delta t) \rangle \delta_{xx'}.
\end{align}
{
	Alternatively, one can work in the time-frequency domain using the
	short-time Fourier transforms of the strains $s^{x}(t)$ and $s^{x'}(t)$
	and substituting in Eq.~\eqref{eq:cc_definition}. The timescale 
	used to compute the short-time Fourier transform should not be greater
	than the timescale of cross-correlation $\tau$.
}
In the remaining analysis, for concreteness, we will work in the time domain. 

\subsection{Choice of residual Templates}\label{sec:formalism_basis}

The second ``key-physics"-driven aspects that we want to explore in CRPS
signals is that there might be more energy injection/extraction from a binary system
in BM scenarios than in the SM templates. This difference in power released
results, in particular, in a change in the orbital frequency evolution $f(t)$. 
{Previous residual tests, such
as the ones using
\textsc{BayesWave}~\cite{cornishBayeswaveBayesianInference2015,cornishBayesWaveAnalysisPipeline2021}, have
proposed to decompose residual signals in terms of frames that have a natural
expression in the frequency domain (Morlet-Gabor sine-Gaussian wavelets and
``chirplets"~\cite{cornishBayeswaveBayesianInference2015}).}
Along similar arguments, we propose
to decompose the CRPS signal into a template of $n$ orbital frequency-dependent
functions $Z_i(f(t))$, {which can capture any gain/loss in the (radiated) energy of the
binary source after taking into account the dependence on the source parameters}
\begin{equation}\label{eq:cc_expansion}
    S^{x x'}_{\theta_\text{MLE}}(t) 
    =
    \sum_{i=1}^{i=n} \alpha_{i}(\theta_\text{MLE},t) Z_i(f(t)), 
\end{equation}
where the (real) coefficients $\alpha_i$ may depend on the
source parameters and time. These coefficients vanish when no BM signature is
present in the data.  Either side of the equation is valued in
strain squared and will depend on the integration timescale $\tau$.
The only requirement on the set of templates $Z_{i}(f(t))$ is that its elements
should not be parallel to one another. They need not be either normalized or
orthogonal. For example, one could choose wavelets or chirplets.
Notice that one can also model the BM signature in terms of the residual
strain of the GW signal rather than the power. As one still uses the CRPS to
extract posterior information in either approach, they will yield the same
information. We describe the strain modeling in Appendix~\ref{app:strain}.

The specific form of the template functions $Z_{i}(f(t))$ can be motivated by a
particular type of BM signature if such knowledge is available. As an
alternative to choosing a specific, model-dependent set of template functions,
we present a choice of template that we argue allows for an unmodeled BM
search. The unmodeled template is constructed from the following observation:
BM effects modify the rate of energy dissipation. This in turn impacts the rate
at which orbital frequency increases. We thus expect the derivatives of the
orbital frequency of the system to be good tracers of a BM signature. 

More specifically, to capture the proportional change in time of the
orbital frequency, we use the derivatives of $\dot{f}/f$ as our
template functions
\begin{align}
	Z_{i>0}(t)
	&=
	\left\langle
		\left( 
			\frac{d^i \ln f(t)}{dt^i}
		\right)^{2}
	\right\rangle,
	\label{eq:toy_model_expansion}
\end{align}
while $Z_{0}(t)$ is constant, defined to capture any BM signature with constant
power. 
The template functions take values at bin positions. If the template functions
are defined as functions of continuous time, one can simply take the angular
bracket, as defined in Eq.~\eqref{eq:cc_definition}, to define the expected
CRPS functions.

We note that for inferring the values of $\alpha_i$ from data, it is convenient to
orthogonalize and normalize the template functions, e.g. using the Gram-Schmidt
process, if possible. If the physical interpretation of the coefficients
is easier in the original form, then one can always transform the functions and
their coefficients back.

\subsection{Projection {on a template}}

{
The projection of the CRPS of the data onto the chosen template encodes how
well the model matches the signal. In this subsection, we define the projection
coefficients and the associated uncertainty. The projection coefficient $\Gamma_{i}(D)$ of the
CRPS signal $D^{x x'}(t)$ onto the template $Z_{i}$ is: 
}
\begin{align}
	\Gamma_{i}(D)
	&=
	\int^{t_e}_{t_s}
	dt
	D^{x x'}(t)
	Z_{i}(f(t))
	,
\end{align}
($t_s$, $t_e$ are the start/end of the signal in the data). Then,
the requirement that the template functions should not be parallel is
{
	\begin{align}
		Z_i - \Gamma_{i}(Z_j) \tilde{Z}_i \neq 0\, \forall \, \{j \neq
		i\} \in n,
	\end{align}
where $\tilde{Z}_i$ is normalized, so $\Gamma_i(\tilde{Z}_i)=1$.
}
The templates can be made orthonormal, that is $\Gamma_{i}(Z_{i})=1$ and
$\Gamma_{i}(Z_{j})=\delta_{ij}$, if required for the inference of the signal.
If an orthonormal set is chosen, then the projection coefficient
$\Gamma_{i}(S)$ will be equal to $\alpha_{i}$. 
This is possible as long as the template functions are not linearly dependent.
In the rest of this work, we will assume that the $Z_{i}$ used
are orthonormal and hence the value of $\alpha_{i}$ is measured as the mean value of the projection of the data onto the
template functions.

{When measuring the CRPS and the coefficients, we will need to
know the associated uncertainty. In this work, we compute the standard deviation in the limit of small-signal (zero mean value). }
In the limit where the cross-correlation scale goes to zero, $\tau \rightarrow 0$, 
the noise on the cross-correlation estimator can be obtained as
\begin{equation}\label{noisecc}
    K^2_{n} 
    \equiv
    \langle (\Delta d^x(t) \Delta d^{x'}(t+\Delta t))^2 \rangle  
    -
    D(t)^2.
\end{equation}
Assuming that the noise in each detector is Gaussian, stationary, and
uncorrelated, we can simplify the above expression as\footnote{The angular
brackets are defined in Eq. \eqref{eq:cc_definition}.}
\begin{align}\label{noisecc1}
\begin{split}
    K^2_{n} \equiv& \langle n^x(t)n^x(t) \rangle \langle n^{x'}(t)n^{x'}(t) \rangle,\\
    =& N^{2}_{x}N^{2}_{x'},
    \end{split}
\end{align}
where in the second line we have defined the noise auto-correlation of each
detector as $N^{2}_{x}$. 

If there are no BM signatures present in the data, then the value of
the coefficients  $\alpha_i$ should be consistent with zero. However, in the
case of a non-zero signal,  we can write the estimator $\hat
\alpha_i$ from the data. Assuming $\alpha_{i}$ does not depend on time, 
then we can write it in terms of the CRPS $D(t)$, the template
functions  $Z_{i}(f(t))$, and the noise on the CRPS, $K_{n}$, as

\begin{align}\label{eqprjc1}
\begin{split}
    \hat \alpha^{xx'}_i (\theta_\text{MLE}) 
    &= 
    \int_{t_s}^{t_e} dt 
    W_i(t) 
    D^{xx'}(t)
    Z_{i}(f(t)),\\
    \text{where\,} W_{i}(t)
    &\equiv
    \frac{K^{-2}_n(t)}
    { \int_{t_s}^{t_e} K^{-2}_n(t) Z_{i}^2(f(t)) dt}.
    \end{split}
\end{align}
In the above equation, $W_{i}(t)$ are weights which serve two purposes: (i)
they ensure that the equation is normalized, and (ii) they account for any
variation in the noise properties in the detector by inverse noise weighting.
This helps in reducing the uncertainty in the estimator. For the situation of
uncorrelated stationary Gaussian noise considered in this analysis, the values
of $W_{i}(t)$ will be constant and will simply normalize $Z_{i}(t)$. However if
the noise {power spectral density} (PSD) across a data segment shows variation
with time $N_x^2(t)$ (non-stationarity), then $W_{i}(t)$ will not be a constant.
{If the $Z_i(f(t))$ are not orthogonal, $\hat{\alpha}_{i}$ will generally be
	non-zero even when $D(t)=Z_j(t)$, for $j\neq i$.}

Finally, it is of interest to estimate the signal-to-noise ratio (SNR) for
measuring the parameters $\hat \alpha$.  For $N_{GW}$ sources with GW similar
source parameters and with $N_{\rm det}$ GW detectors, 
we can write the
combined SNR, $\rho_i$, as 
\begin{equation}\label{totsnr}
   \rho_i
   =
     \bar{\alpha}_i (\theta_\text{MLE})\times
	\bigg[\sum^{N_{\rm det}}_{x=1, x'>x} \sum^{N_{GW}}_j 
    {{(K^{x x'}_{\alpha_{i}})}^{-2}}\bigg]^{1/2}
    ,
\end{equation}
where, $\bar{\alpha}_i (\theta_\text{MLE})$ is the mean value of the inferred
signal for all the detectors pairs and sources, and {$(K^{x
x'}_{\alpha_{i}})^2$} is the noise on the parameter $\alpha_i$, defined as
\begin{equation}
   (K^{xx'}_{\alpha_{i}})^2
   = 
	   \int_{t_s}^{t_e} dt 
	   \left(
	   W(t)
	   K_{n}
	   Z_{i}(t)
	   \right)^{2}
	   .
    \label{eq:sigma_2_alpha}
\end{equation}

From the  expression given in Eq. \ref{totsnr}, we can conclude that the total SNR $\rho$ to measure a deviation $\alpha_i$ scales as $\sqrt{N_{GW}}$ and $\sqrt{N_{\rm det}(N_{\rm det} -1)/2}$.
Finally it is important to clarify that the value of $\alpha$ can
explicitly depend on the value of the source parameters $\{\theta_\text{MLE}
\}$, and in that case so will the number of GW sources
$N_{GW}(\theta_\text{MLE} )$ with the best-fit source parameter
$\theta_\text{MLE} $ that needs to be combined to obtain the SNR.  

\subsection{{Inference using the Bayesian framework}}\label{Bayesian_framework}

In this section we formulate a hierarchical Bayesian framework to measure the
presence of any BM signature in the observed data by
marginalizing over the uncertainties associated with the GW source parameters.
Let us denote $\mathcal{M}_\text{SM} $ as the SM, and $\Delta_M$ as a set of parameters which
captures deviations from this model. Then, we can write the posterior
distribution $\mathcal{P}(\Delta_M|\{d^x(t)\})$ on the parameters $\Delta_M$
given the observed set of data $\{d^x(t)\}$ detected at detectors labelled
$x$ using Bayes' theorem~\cite{bayesLIIEssaySolving1763} as
\begin{equation}\label{bayes1}
    \mathcal{P}(\Delta_M|\{d^x(t)\})\propto p(\{d^x(t)\}|\Delta_M)\Pi(\Delta_M),
\end{equation}
where $p(\{d^x(t)\}|\Delta_M)$ denotes the likelihood and $\Pi(\Delta_M)$ denotes the prior. For a set of $N_{\rm obs}$ independent events of GW sources detected (denoted by $\mathcal{S}$) above a matched filtering SNR $\rho_*$, we can simplify the likelihood as
\begin{align}\label{bayes2}
\begin{split}
    &p(\{d^x(t)\}|\Delta_M)\\
    =&
    \prod^{N_{\rm det}}_{x,x'>x=1}
    \prod_{i=1}^{N_{\rm obs}}
    p(\{d^x, d^{x'}\}_i |\mathcal{S},\Delta_M)\\
    =&
    \prod^{N_{\rm det}}_{x,x'>x=1}
    \prod_{i=1}^{N_{\rm obs}}
    \frac{p(\mathcal{S}| \{d^x, d^{x'}\}_i,\Delta_M)
    p(\{d^x, d^{x'}\}_i|\Delta_M)}
    {p(\mathcal{S}|\Delta_M)},
    \end{split}
\end{align}
where in the third line, the first term denotes the probability of detecting
events in the data given a small variation in the model $\Delta_M$, and the second
term $p(\{d^x, d^{x'}\}_i|\Delta_M)$ is the likelihood. We can write the
likelihood including the GW source parameters $\theta$ with a prior
$\Pi(\theta)$, and the dependence of any deviation on the GW source parameter, 
$p(\Delta_M|\theta)$, as
\begin{equation}\label{bayes3}
	p(\{d^x, d^{x'}\}_i|\Delta_M)
    =
    \int d \theta p(\{d^x, d^{x'}\}_i|\theta, \Delta_M)
    p(\Delta_M|\theta)\Pi(\theta).
    \end{equation}
The term in the denominator in Eq. \ref{bayes2} is the evidence, which we can
write in terms of the all possible events which are detectable above a matched
filtering SNR $\rho_*$ as 
\begin{align}\label{bayes4}
    &p(\mathcal{S}|\Delta_M) \nonumber \\
    &= \int_{\rho\geq \rho_*} d\{d^{x}(t)\} \int d \theta
    p(\{d^{x}(t)\}|\theta, \Delta_M) p(\Delta_M|\theta)\Pi(\theta).
\end{align}
If the detector noise is not changing with time, then the term
$p(\{d^{x}(t)\}|\theta, \Delta_M)$ will not vary with time and only depend on the
detectability of an event above $\rho_*$, which we usually denote by $p_{\rm
det} (\theta, \Delta_M)$. If we assume, any signal with a matched filtering
SNR above $\rho_*$ is detected, then Eq.~\eqref{bayes2} further simplifies. The
term $\int p(\mathcal{S}| \{d^x, d^{x'}\}_i,\Delta_M)$ becomes unity. Putting all
of this together, we can write the posterior on the parameters $\Delta_M$ as
\begin{align}\label{bayes5}
\begin{split}
    \mathcal{P}(\Delta_M|&\{d^{x}(t)\})
    =
    \Pi(\Delta_M)
    \times \\
    &\prod^{N_{\rm det}}_{x,y>x=1}
    \prod_{i=1}^{N_{\rm obs}}
    \frac{
	    \int
	    p(\{d^x, d^{x'}\}_i|\theta, \Delta_M) 
	    p(\Delta_M|\theta)
	    \Pi(\theta)d\theta
	}
    {p(\mathcal{S}|\Delta_M)}.
    \end{split}
\end{align} 
This is the most general hierarchical Bayesian framework to search
for a BM signature marginalizing over the source parameters uncertainties. In this analysis, we perform the Bayesian estimation on the residual signal obtained
around a set of  best-fit model parameters $\theta_{\rm BF}$ assuming
$\mathcal{M}_{SM}$ as the SM.
{
	This allows us to define the deviation $\Delta_{M}$ as a function of
	$\theta_\text{BF}$ so we may check any correlation between BM
	signature and source parameters. Searches around the best-fit can also help in
reducing the computational cost in performing the Bayesian analysis.} However, the analysis pipeline can be easily modified to include the full Bayesian framework. The above equation around a best-fit value
can be simplified into 
\begin{widetext}
\begin{align}\label{bayes6}
%\begin{equation}
    \mathcal{P}(\alpha|&\{d(t)\})
    =
    \Pi(\alpha) 
    \prod^{N_{\rm det}}_{x=1,y>x}
    \prod_{i=1}^{N_{\rm obs}}
    \frac{
	 \int
	 p(\{d^x, d^{x'}\}_i|s(\theta_{\rm MLE}), \alpha) 
	p(\alpha|\theta_{\rm  MLE})
	p(\theta_{\rm MLE})
	d \theta_\text{MLE} 
    }
    {p(\mathcal{S}|\alpha)},
    %\end{equation}
\end{align}
\end{widetext}
where, $s(\theta_{\rm BF})$ is the GW signal for the set of best-fit
parameters, $p(\theta_{\rm BF})$ is the posterior on the best-fit parameters
obtained from the data assuming the SM, and $\alpha$ is the set of
template coefficients for $\Delta_{M}$.  The
likelihood $p(\{d^x, d^{x'}\}_i|s(\theta_{\rm BF}), \alpha)$ can be written in
terms of the CRPS as 
\begin{align}\label{bayes7}
\begin{split}
	p(\{d^x, d^{x'}\}_i|s(\theta_{\rm BF}), \alpha)= p(D(t)|\alpha).
    \end{split}
\end{align}
If we further assume the likelihood to be Gaussian and there is no correlated noise, then it becomes
\begin{align}\label{bayes8}
\begin{split}
	&	p(\{d^x, d^{x'}\}_i|s(\theta_{\rm BF}), \alpha)\\
	 &\propto
   \exp\bigg(-\int dt \frac{(D(t) - S_{\theta_\text{MLE}} (\alpha))^2}{2K_n^{2}(t)}\bigg),
\end{split}
\end{align}
where $D(t)$ can be calculated using Eq. \eqref{cc1} and the noise covariance matrix can be calculated using Eq. \eqref{noisecc}. 
In the presence of correlated noise $\mathcal{N}_c$ it can be included in the noise covariance matrix. 

{
Though this method can detect the presence of any kind of BM signature in the GW data,
model-independent searches may fail to {completely} capture the
structure of the deviation if the chosen template only partially overlaps with
the BM signature. However, using the Bayesian framework proposed here, one can
do a Bayesian model comparison for different BM scenarios and perform a
tuned search for the model with higher Bayesian evidence. }

\section{Application of SCoRe on a toy example for an unmodeled search}\label{results_1}

We now illustrate the \texttt{SCoRe} method by using the unmodeled power
template to recover a toy model injection from simulated data. We describe the
toy model, then discuss the appropriate choice of cross-correlation timescale.
Using the \texttt{SCoRe} framework, we finally recover the injected value in
the simulated data and perform the Bayesian analysis described in
Sec.~\ref{Bayesian_framework}.

\subsection{Generating mock data}\label{sec:toy_model_data}

As a toy model on which to apply this template, we create simulated mock data
by adding Gaussian noise realisations $n^{x}$ over a simulated strain $s(t)$ as
\begin{equation}\label{eq:toy_model_strain}
	d^{x}(t) = s(t)	+	n^{x}(t).
\end{equation}
The strain is $s(t)\equiv s_\text{SM} (t) + \Delta s_\text{BM}$; it includes a
standard model signal $s_\text{SM} (t)$ and a BM signal $\Delta s_\text{BM}$.
The BM signal is modeled as $\Delta s_\text{BM} (t)= b \frac{d \ln
f_\text{SM} (t)}{d t}$, with $b$ as a free parameter that controls the strength
of the BM signal. The standard model waveform is the same for all events. It is
computed  using LALSuite~\cite{lalsuite} called through PyCBC~\cite{nitzGwastroPycbcV22022} and the IMRPhenomD approximant~\cite{husaFrequencydomainGravitationalWaves2016,
khanFrequencydomainGravitationalWaves2016} for an equal mass, non-spinning,
circular BBH merger, at a luminosity distance of $100$ Mpc
and the individual masses are both set to $m=5M_{\odot}$.
A sampling rate of $16384$Hz is used throughout this work. Both cross-
and plus-polarizations are used to derive $f_\text{SM}$. The
plus-polarization is used as the standard model signal $s_\text{SM}$. We
assume Gaussian, stationary noise and the same sensitivity for all
detectors. Different events in each of the two distinct detectors are
then simulated by drawing realisations $n^{x}$ of the expected O4 noise
sensitivity~\cite{abbottProspectsObservingLocalizing2020}.\footnote{We have used
the noise file aLIGOAdVO4T1800545 in PyCBC.}

We project the CRPS onto the unmodeled template given
in~\eqref{eq:toy_model_expansion}.
In the rest of the work, we implement the
angular bracket average of Eq.~\eqref{eq:cc_definition} as a binned
mean.\footnote{One could also use a running mean, instead of a binned mean
value.} Out of the template terms given in Eq.~\eqref{eq:toy_model_expansion},
we only use the linear term $Z_{1}$. \footnote{ Since we are only using one
	template function, our template is orthogonal by construction.
}
The toy model with the BM signal $\Delta
s_\text{BM}$ is constructed such that it perfectly overlaps with the template 
$Z_{1}$ to show how the recovery can happen for the best case scenario. 
We expect that the template coefficient $\alpha_{1}$ can recover the injected BM
signal $\Delta s_\text{BM}$ completely. Certainly, in reality different BM
theories will only have a partial projection on the template, unless one performs
a model-dependent search.  In the future, we will explore different choices of
template to illustrate how they can project onto different BM theories.  Theoretical
efforts to produce waveforms in some such theories have been presented in
e.g.~\cite{Barausse:2012da,Bernard:2018hta} and so have their impact data
analysis in modeled and unmodeled searches
e.g.~\cite{Sampson:2014qqa,Edelman_2021}.

\begin{figure}
	\includegraphics{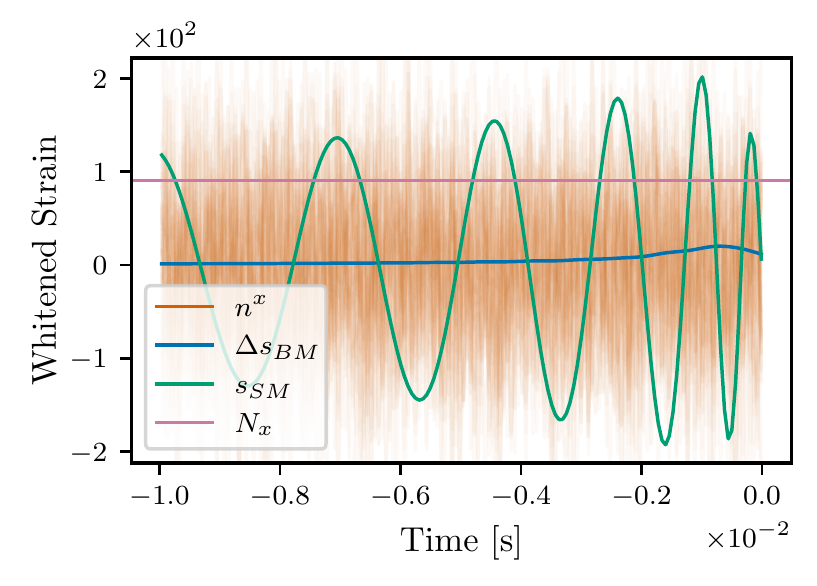}
	\caption{Example toy model data. We use the PyCBC
software package to generate $s_\text{SM}$,
the waveform for a
circular, non-spinning, equal mass BBH,  with individual masses
both equal to $5M_{\odot}$.
This is plotted in
green. We add onto it a
BM signature, $\Delta s_\text{BM}$ (blue line), that is proportional
to the change in the orbital frequency logarithm ($\tilde{\alpha}_1=0.05$). 
It is normalized so that, for $\tilde{\alpha}_1=1$, the maximum amplitude reaches to the noise
auto-correlation, $N_x$ (purple line). Realisations of Gaussian, stationary noise are added to $s_\text{SM} + \Delta s_\text{BM}$ to obtain different events. Some examples of noise realisations are plotted in orange.}.
	\label{fig:strain}
\end{figure}

We generate $N_\text{GW} $ pairs of strain data as in
Eq.~\eqref{eq:toy_model_strain} and attribute each of these strains to one of the 
two detectors $x$ or $x'$. Examples of each of the components used to construct
these data are shown in Fig.~\ref{fig:strain}. We assume that the best fit parameters are known exactly, and that they are
those used to generate $s_\text{SM} (t)$. We therefore subtract $s_\text{SM}$
from each of these strains to obtain $\Delta d^{x,x'}$. In real data, the error
from estimating $\theta_\text{MLE}$ needs to be accounted for both in the
residual strains and in the template and we need to perform the marginalization
described in Sec.~\ref{Bayesian_framework}.

\begin{figure}
	\includegraphics{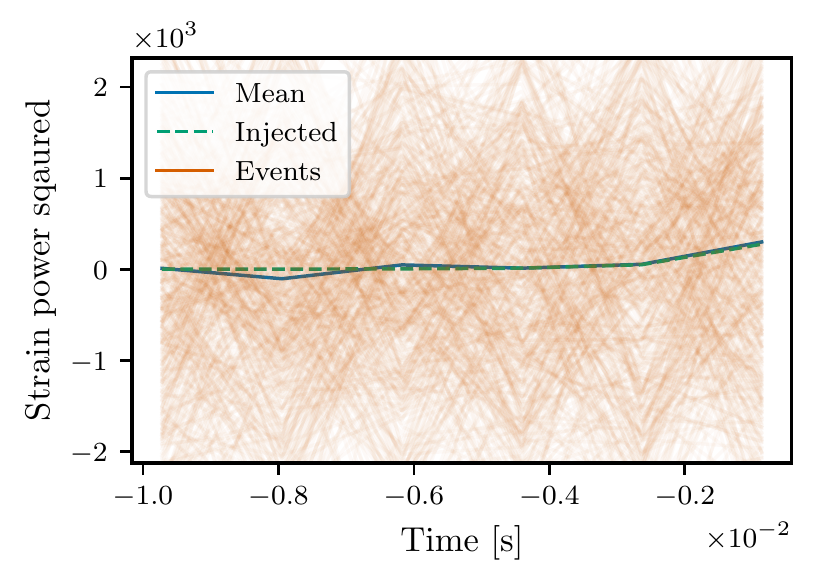}
	\caption{Residual data and injected BM signature cross-correlated over $\tau =
		1.28 \times  10^{-3}$ s. The blue 
		line is the mean of the CRPS over 500 events, plotted in orange.
		It corresponds to the MLE value of $\tilde{\alpha}_1$ when
		projected onto the template. The dashed green line
	is the cross-correlation of $\Delta s_\text{BM}$. }
	\label{fig:crosscorrelation}
\end{figure}

Each pair of $\Delta d^{x,x'}$ is then cross-correlated as in
Eq.~\eqref{eq:D_definition} to obtain the CRPS, $D^{x,x'}_i$, where $i$ labels 
events. We obtain $N_\text{GW}$ residual time series that, for real data, would
each be associated with a GW event. Examples of the CRPS 
are shown by orange lines in Fig.~\ref{fig:crosscorrelation}. An
estimator $\hat{\alpha}_1$ is obtained for each of these time-series by
projecting onto the template according to Eq.~\eqref{eqprjc1}. It is convenient
to define the dimensionless quantities
\begin{align}
\begin{split}
	\tilde{Z}_{1}	
	&=
	P^{-1} Z_{1}, \quad
	\text{such that}
	\int_{t_\text{s}}^{t_\text{e} }
	\tilde{Z}_{1}^{2} dt
	=
	1,\\
	\tilde{\alpha}_{1}
	&=
	\frac{\alpha_{1}}
	{\alpha ^{0}_{1}},
	\quad
	\alpha_{1}^{0}
	=
	\frac{P N_{x}^{2}}
	{
		\text{max}\left( \frac{d \ln f}{dt}  \right) ^{2} 
	},
	\end{split}
\end{align}
where $\alpha ^{0}_{1}$ is defined so that, when the maximum
amplitude of $\Delta s_\text{BM} $ is $N_{x}$, the projection of its
cross-correlation  onto $\tilde{Z}_{1}$ is $\tilde{\alpha}_{1}=1$. The normalization constant, $P$, is a measure of the mean power per bin in the template function. During this
procedure, the timescale $\tau$ over which the residual strains are
cross-correlated is a free parameter. We discuss the choice of $\tau$ in the next subsection.

\begin{figure}
	\includegraphics{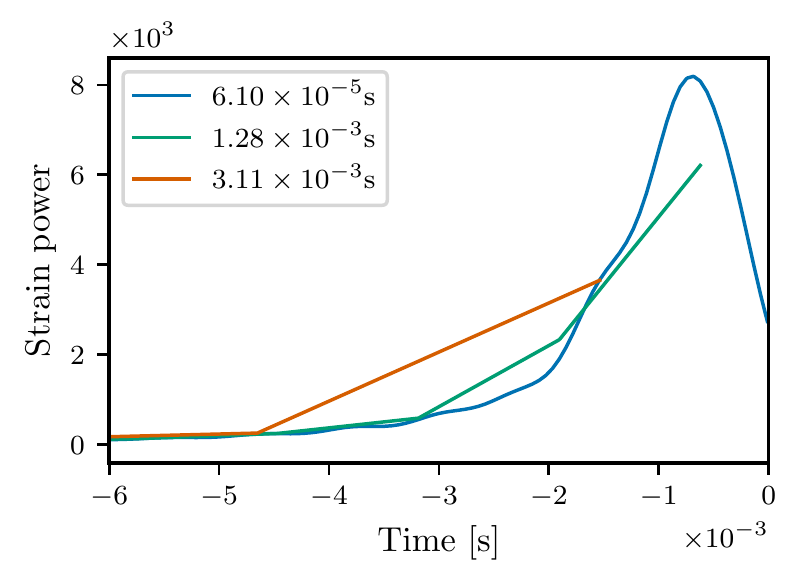}
	\caption{Cross-correlation template $\tilde{Z}_1$ for different
	cross-correlation timescales $\tau$. The smallest timescale allowed by
	the sampling rate, $\tau=6.10\times 10^{-5}$ s, has features that are
	washed away as the timescale is increased. With $\tau \in [0.8,1.6]
	\times 10^{-3}s$, no new physical features appear and information
	remains the same as $\tau$ is increased.
	For timescales longer than $2.1 \times
	10^{-3}$ s, all the power is effectively concentrated in a single bin.}
	\label{fig:bases_tau}
\end{figure}

\subsection{Choice of $\tau$}\label{sec:choice_of_tau}

When $\tau$ is finite, the noise on the cross-correlation estimator, in the
limit of a small signal is
\begin{align}
	K^{2}_D	
	&=
	\frac{1}{n_\text{window} }
	K^{2}_{n},
\end{align}
where $n_\text{window}$ is the number of data points measured in the timescale
$\tau$. The advantage of taking the mean over a timescale $\tau$ is that $K_{D}
\propto 1/\tau$, so, in the limit of large $\tau$, the cross-correlation tends
to be the residual signal (and correlated noise). On the other hand, as
mentioned, some information is lost in the averaging {over the timescale
$\tau$}.  This is illustrated in Fig.~\ref{fig:bases_tau}. The
cross-correlation template $\tilde{Z}_1$ is computed from $s_\text{SM}$ with
different values of $\tau$. As $\tau$ is increased, the shape of
$\tilde{Z}_{1}$ is qualitatively changed. By integrating over $\tau$, we lose
the ability to differentiate between BM models that give different template
predictions below this timescale.

\begin{figure}[h]
	
	\includegraphics{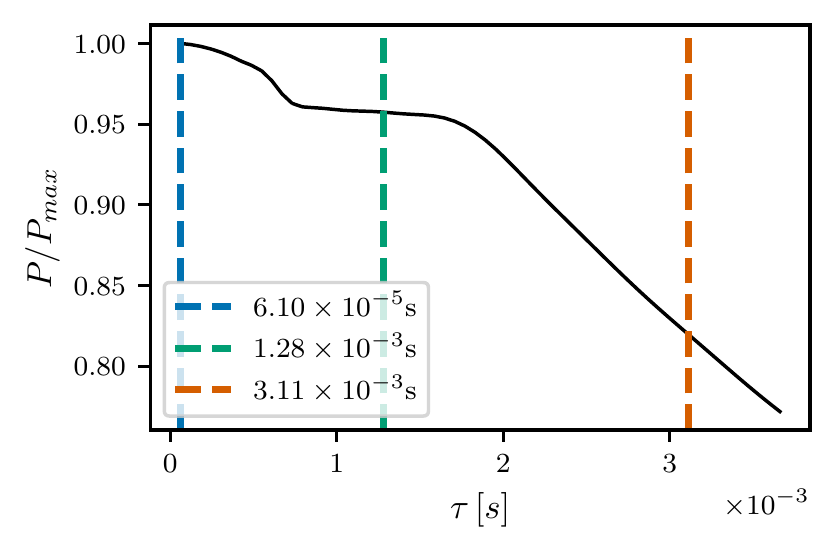}
	\caption{Template normalization factor (mean power per bin $P$) as $\tau$ is increased. The template
	with $\tau = 6.10 \times 10^{-5}$ s contains a pulse. As the pulse is
	washed away by the averaging, the mean power per bin decreases. Between 
	$0.8 \times 10^{-3}s < \tau < 1.6 \times 10^{-3}s$, averaging over a
	greater $\tau$ does not change the shape of the template and the mean power
	remains constant. At scales greater than $2.1 \times 10^{-3}$ s a single
	bin contains the majority of the power and the mean power scales as
	$\tau^{-1/2}$.}
	\label{fig:alpha_vs_tau}
\end{figure}

For a specific template, two choices of $\tau$ have equivalent signal content if
no new physical behavior arises at an intermediate timescale. If they are
equivalent, the average power per bin, $P$, will be the same. 
This quantity is plotted in Fig.~\ref{fig:alpha_vs_tau}
as a function of $\tau$. {For this example,} at timescales shorter than $0.8 \times 10^{-3}$ s, the
mean power decreases as the ``pulse" feature of the template seen in
Fig.~\ref{fig:bases_tau} is integrated out. For $\tau \in [0.8,1.6] \times
10^{-3}$ s, the mean power remains constant as the template functions resemble power laws and
a smaller $\tau$ does not add information. An example of a template with a
timescale in this range is shown by the green line in Fig.~\ref{fig:bases_tau}
(obtained by averaging over $\tau = 1.28 \times 10^{-3}$ s). When averaging
over timescales larger than $\tau \sim 2.1 \times 10^{-3}$ s, all the power of
the BM signature is concentrated in a single bin. Past this point, the mean
power per bin, $P$ scales as $\left( \tau \right) ^{-1/2}$.

This exact evolution sequence of the mean power per bin $P$ as a function of
$\tau$ is specific to our chosen template~\eqref{eq:toy_model_expansion} and
the standard model waveform used for the computation. However, the criteria
that can be used to choose $\tau$ will be the same in all cases.

The first natural choice of $\tau$ corresponds to the maximum SNR, which is
achieved when most of the power is concentrated in as small a bin as possible.
This is determined by the $\tau$ where the mean power per bin, $P$,
starts scaling as $\tau^{-1 / 2}$. This may not, however, correspond to the
highest Bayes factor in favor of a BM.

If information about the structure of the BM signal is preferred over a higher
SNR, then $\tau$ should be accordingly chosen. In our toy model, for our
specific choice of template and SM waveform, to
conserve information about the ``chirp" part of the template, the $\tau$ value
needs to be set to the rightmost point on the horizontal section of the  $P$
shown in Fig. \ref{fig:alpha_vs_tau}, $\tau=1.8 \times 10^{-3}$ s. 
{
In this case, 
it is not possible to preserve information about the pulse shape while still
taking the cross-correlation, as one would need to set $\tau$ to the sampling
interval.
It may yet be possible to keep some benefits of taking the cross-correlation
while not averaging over the pulse shape by letting the averaging timescale
vary with time. For example, we may set $\tau$ to 1 over the sampling rate from
$t=-10^{-3}$ s to $t=0$, and set $\tau \in (0.8,1,6) \times 10^{-3}$ s
for $t \in (-5,-1) \times 10^{-3}$ s to maximise noise reduction over the
``chirp".
}
Finally, we could average the rest of the time domain, where the CRPS
is dominated by noise, into one bin ( $t=(t_\text{s},-5) \times 10^{-3}$ s). 

In this section, we consider only a time-constant $\tau$, which we set to the
SNR-maximising value, $\tau=1.8 \times 10^{-3}$ s.  The cross-correlations
plotted in Fig.~\ref{fig:crosscorrelation} are computed over this timescale. 
{
This choice of $\tau$ is specific to this template function and
best-fit source parameters, as the variation of the mean power $P$ with $\tau$
will change for other choices of template and source parameters. Since the
templates depend on the orbital frequency,
we expect the choice of $\tau$ to vary (scale) with chirp mass for all choices of templates.
}

\subsection{Bayesian inference of the injected
signal}\label{sec:results_1:toy_model}

In this section, we use the Bayesian framework presented in
Sec.~\ref{Bayesian_framework} to combine events into one measurement of
$\tilde{\alpha}_1$. We assume the source parameters are perfectly known to be
$\tilde{\theta}_\text{BF}$. Therefore, $p ( \tilde{\theta}_\text{BF} ) = 
\delta ( \theta_\text{MLE} -\tilde{\theta}_\text{BF}   )$, where
$\delta(x)$ is the Dirac delta function. This assumption is made to show how
this new method will work in the best-case scenarios. In reality, each
estimated parameter will have an error, and one can marginalize the uncertainty
as presented in Sec.~\ref{Bayesian_framework}. Furthermore, since we assume
$\alpha_1$ is unique for the given source parameters, $p(
\alpha_{1}|\tilde{\theta}_\text{BF}  )=1$. The posterior on
$\tilde{\alpha}_{1}$ is then
\begin{align}  \label{eq:toy_model_posterior}
\begin{split}
  \mathcal{P}
  &
  \left( \tilde{\alpha}_{1} | \{d(t)\} \right)
  \\
  &\propto
  \Pi(\tilde{\alpha}_{1})
  \exp
  \left( 
	  - \frac{1}{2 \tilde{K}^{2}_{\tilde{\alpha}}}
	  \left( 
		  \tilde{\alpha}_{1} 
		  -
		  \tilde{\alpha}_{1,MLE}
	  \right)^{2}
  \right),
  \end{split}
  \end{align}
where $\tilde{\alpha}_{1,MLE}
	=
	1/(N_\text{GW} \alpha_{1}^{0} )
	  \sum^{N_\text{GW} }_i
	  \int^{t_\text{e} }_{t_\text{s} } \tilde{Z}_{1} D_{i} dt$,
and 
\begin{equation}
\tilde{K}_{\tilde{\alpha}_1}^{2}
=
\frac{K^{2}_{\alpha_{1}}}{\left( a_{1}^{0} \right)^{2} N_\text{GW} }.
\end{equation}

\begin{figure}
	\includegraphics{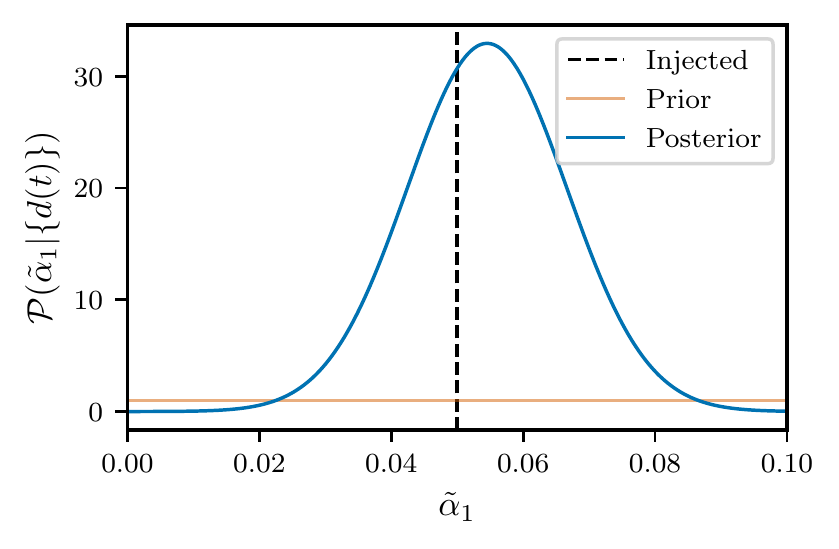}
	\caption{Posterior on $\tilde{\alpha}_{1}$ for $N_\text{GW}=500$,
		when the template function $Z_{1}$ is injected in Gaussian,
		stationary noise.
		We used a flat prior from $0$ to $1$, giving a
		Bayes factor of $773$. The error on $\tilde{\alpha}$ is
		$\tilde{K}_{\tilde{\alpha}}$,
	the SNR for this measurement is 4.5. 
}
		\label{fig:toy_model_posterior}
\end{figure}

The likelihood is thus a Gaussian distribution with
Maximum Likelihood Estimator (MLE) $\tilde{\alpha}_{1,\text{MLE}}$ and variance
$\tilde{K}^{2}_{\tilde{\alpha}_1}$. With our assumption, the MLE is the mean over events
of the estimator $\hat{\alpha}_1$. Since the projection is linear, it
corresponds to the projection of the mean CRPS,
shown in blue in Fig.~\ref{fig:crosscorrelation}. 

\begin{figure*}
	\begin{minipage}{0.49\textwidth}
		\includegraphics{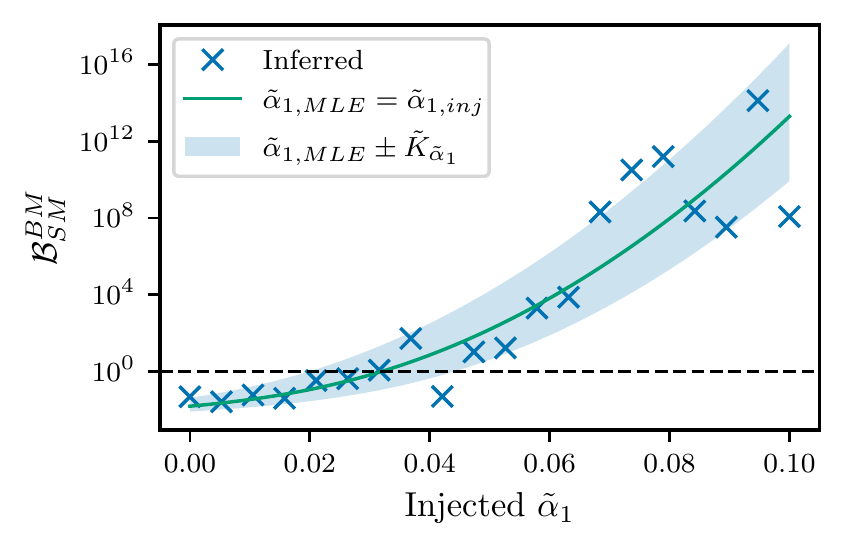}
	\end{minipage}
	\hfill
	\begin{minipage}{0.49\textwidth}
		\includegraphics{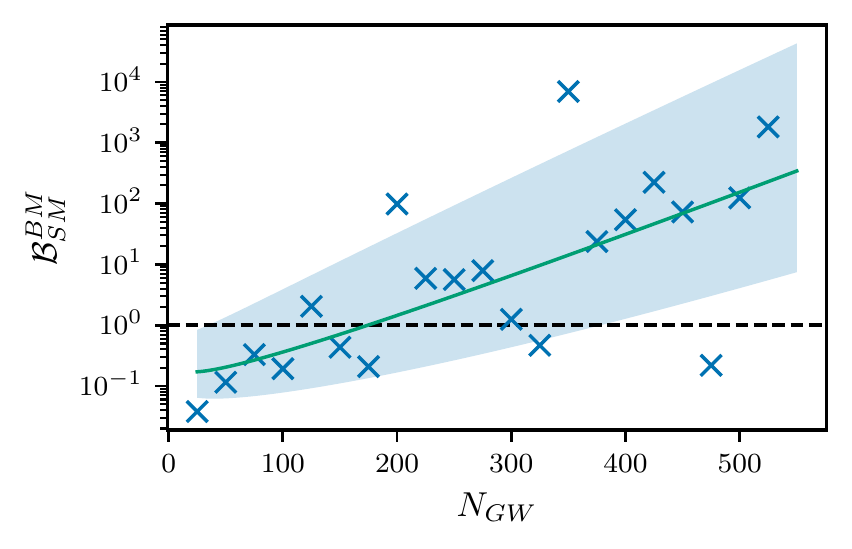}
	\end{minipage}
	\caption{Bayes factor $\mathcal{B}^\text{BM}_\text{SM}$ in favor of a BM
		hypothesis over the standard model hypothesis \textit{left} for
		different values of $\tilde{\alpha}_{1,\text{inj}}$
		($N_\text{GW} = 500$), \textit{right} for different
		$N_\text{GW}$ ($\tilde{\alpha}_{1,\text{inj}}=5\times 10^{-2}$).
		The solid green line indicates the case of perfect
		accuracy as a guide. The blue region covers one
		$\tilde{K}_{\alpha}$ deviations in the measured value of
		$\tilde{\alpha}_\text{1,MLE}$. 
		{ 
		The standard model waveform used is that of an equal mass BBH
		with individual mass $5 M_{\odot}$.
		}
	}
	\label{fig:toy_model_bayes}
\end{figure*}

If a signal is correlated between detectors, its cross-correlation {$S^{x x'}(t)$} must be
positive. Our template function is positive, therefore so must be $\alpha_{1}$
and $\tilde{\alpha}_{1}$. We choose a flat prior on $\tilde{\alpha}_{1}$
ranging from 0 to $\tilde{\alpha}_{1,max}$, so that
$\Pi(\tilde{\alpha}_{1})=\left( \tilde{\alpha}_{1,max} \right)^{-1}$. 
The posterior obtained from the events in Fig.~\ref{fig:crosscorrelation}, 
which have injection $\tilde{\alpha}_1=0.05$, is  
shown is shown in Fig.~\ref{fig:toy_model_posterior}. The prior used is set
with $\tilde{\alpha}_{1,max}=1$.

With this posterior, we can compare the standard model hypothesis with the hypothesis
of a BM signature by computing $\mathcal{B}^\text{BM}_\text{SM}$, the Bayes factor
in favor of a BM signature being present as opposed to just the SM.
Since the standard model is recovered when the coefficient $\alpha_{i}$ is
zero, we compute $\mathcal{B}^\text{BM}_\text{SM}$ as the Savage-Dickey density
\begin{align}
	\mathcal{B}^\text{BM}_\text{SM}
	&=
	\frac{\Pi(0)}{P(\tilde{\alpha}_{1}=0|\{d(t)\})},
\end{align}
which can be computed directly in our toy model, since the analytical form of
the posterior in terms of $\tilde{\alpha}_{1,MLE}$, $K^{2}_{D}$, and
$N_\text{GW}$ is known. In fact, the expression for $\mathcal{B}^\text{BM}_\text{SM}$ can
be written explicitly in terms of error functions
\begin{align}
\begin{split}
	\label{eq:analytical_bayes}
	\mathcal{B}^\text{BM}_\text{SM}	
	&=
	\sqrt{\frac{\pi}{2}}
	\frac{\tilde{K}_{\tilde{\alpha}_1}}{\tilde{\alpha}_{1,\text{max}}}
	\exp \left( 
		-\frac{1}{2}
		\left( 
			\frac{\tilde{\alpha}_{1,\text{MLE}}}{\tilde{K}_{\tilde{\alpha}_1}}
		\right)^{2}
	\right) \times \\
	&\left( 
		\text{erf} \left(
		\frac{\tilde{\alpha}_{1,\text{MLE}}}{ \sqrt{2}\tilde{K}_{\tilde{\alpha}_1}} \right)
		-
		\text{erf} \left(
		\frac{\tilde{\alpha}_{\text{MLE}} -\tilde{\alpha}_{1,\text{max}}}{ \sqrt{2}\tilde{K}_{\tilde{\alpha}_1}} \right)
	\right),
\end{split}
\end{align}
where $	c =
(1/2) \tilde{K}^{2}_{\tilde{\alpha}_1}$ and $\text{erf}(x)= (2/\sqrt{\pi}) \int^x_0 \exp
\left( - y^{2} \right) dy$. This can be computed numerically. We plot the
dependence on $N_\text{GW}$ and on the injected value of $ \tilde{\alpha}_{1}$
in green on Fig.~\ref{fig:toy_model_bayes}, assuming a perfectly accurate
measurement ($\tilde{\alpha}_{1,MLE}=\tilde{\alpha}_{1,\text{inj}}$). In the
left plot, the number of events is kept fixed at $N_\text{GW} =500$ as the
injected value of $\tilde{\alpha}_{1}$ is varied, while on the right plot,
$N_\text{GW}$ is varied as $\tilde{\alpha}_{1,\text{inj}} $ is kept fixed. We
also plot values of $\tilde{\alpha}_{1, MLE}$ measured from realizations of the
toy model data with crosses.

{
The value of $\Pi(0)$ can significantly change the recovered value of the Bayes factor. This
is a manifestation of Occam's penalty~\cite{isiComparingBayesFactors2022}: a
wider prior compared to the likelihood indicates a more constraining underlying
hypothesis. The prior we choose has $\tilde{\alpha}_{1,max}=1$. This maximum
prior is quite large ($1 \gg \tilde{\alpha}_{1}+
\tilde{\Sigma}_{\tilde{\alpha}_1}$), and reducing it can increase the Bayes
factor obtained without truncating most of the area under the likelihood.
However, we choose it on physical grounds, as we assume any yet undetected
BM signature has to be smaller than the noise variance. }

\section{Application of SCoRe on a beyond GR model}\label{results_2}

\subsection{Toy model EFT and numerical set-up}

To explore our method in a realistic scenario --including inspiral, merger and ringdown-- 
we construct a waveform model
that captures key effects expected from an {effective field theory} (EFT) extension of General Relativity under the
assumption that the coupling scale is of the order (or below) that of the 
BH scale.\footnote{If such a scale is larger, as
considered in~\cite{Sennett:2019bpc}, ${\cal O}(1)$ corrections to the GR gravitational
Lagrangian are induced, rendering unclear what describes the regime with smaller
scales (thus including
merger and ringdown), as they would lie outside the EFT regime.}
The model accounts for: (i) higher curvature corrections in the inspiraling regime can
be captured at leading order by tidal effects, (ii) the merger transitions
to a post-merger regime described by quasinormal modes for the black hole
described by such a theory and, importantly, that corrections scale with the
mass of the system appropriately.
To fix ideas, and choose a convenient scaling so its dependence is sufficiently
marked, we envision our system described by a Lagrangian ${\cal L} = R + {l}^{4}{\cal L}_{(6)}$, 
where ${\cal L}_{(6)}$ indicates the higher order corrections to the curvature, scaling in this example as $M^{-4}$ (see e.g.~\cite{DeRham:2018bgz,Cardoso:2018ptl,Cano:2021myl,Cayuso:2020lca}). For modeling
the merger, we follow~\cite{McWilliams:2018ztb} and construct an interpolating signal
between inspiral and ringdown guided by the behavior of null geodesics corresponding to the
final black hole. The main characteristics (mass and spin) of such a final black hole can be estimated
using a strategy similar to that given in~\cite{Buonanno:2007sv,Kesden:2008ga} as long as the general
solution for a rotating black hole in a given theory is known.
For the inspiral regime, we use a simple {effective one-body} EOB model~\cite{Buonanno:2000ef} with
tidal interactions as in~\cite{Hotokezaka_2013} (see also ~\cite{Damour_2012}) transitioning to the final
black hole at the {innermost stable circular orbit} (ISCO) frequency of the final black hole.
For concreteness, we focus on the equal mass, non-rotating, binary black holes with no
eccentricity.~\footnote{As noted in~\cite{Bhat:2022amc} eccentricity can introduce a 
systematic bias in the identification of departures from GR---a potential issue that should be
contemplated carefully in the analysis.}
Importantly, we note that for the theories described by the above Lagrangian, 
closed-form solutions are only known for slowly (or non) rotating scenarios while
the merger outcome should yield a black hole with spin $a/M \approx 0.67$ --with
a small correction that also scales as $M^{-4}$. Without such a solution, the specific
value of such correction can not be computed yet. However, we note that
this correction is quite smaller than the correction to the horizon 
area (e.g.~\cite{Cardoso:2018ptl}) and so black hole compaction is more affected
than associated quasinormal modes. Last, gravitational radiation from this theory can
be captured to leading order by the correction to the Quadrupole of the system; its
associated effect, in turn, can be accounted for in the stationary phase approximation
to the waveform as discussed in~\cite{Sennett:2019bpc}.
Combining the above information, we ``phenomenologically'' 
account for dynamical departures in the theory by extrapolating from the slowly rotating 
case with the following assignments:\\
\begin{eqnarray}
C = \frac{1}{2} - \frac{5}{16}\epsilon \, , \,
k^T = \frac{1008}{25} \epsilon \, , \,  \\
\delta \omega^R_{\mbox{QNM}}/{}^{GR}\omega^R_{\mbox{QNM}} = 1 + 0.45 \epsilon,  \\
\delta \omega^I_{\mbox{QNM}}/{}^{GR}\omega^I_{\mbox{QNM}} = 1 - 2.75 \epsilon,  
\end{eqnarray}
with $C$ the compaction of each black hole, $k^T$ the tidal Love number,
 $\delta \omega$ the correction to QNMs and where
 $\epsilon=(M_*/M_{T})^4$ and $M_*$ a base scale which we take to be $=M_{\odot}=1.5$ km. That is, the EFT is such that it assumes corrections to General Relativity, become ${\cal O}(1)$ at such 
 length, thereby affecting the inspiral only at a subtle level.

\subsection{Choice of template}

\begin{figure*}
	\begin{minipage}{0.48\textwidth}
	\includegraphics{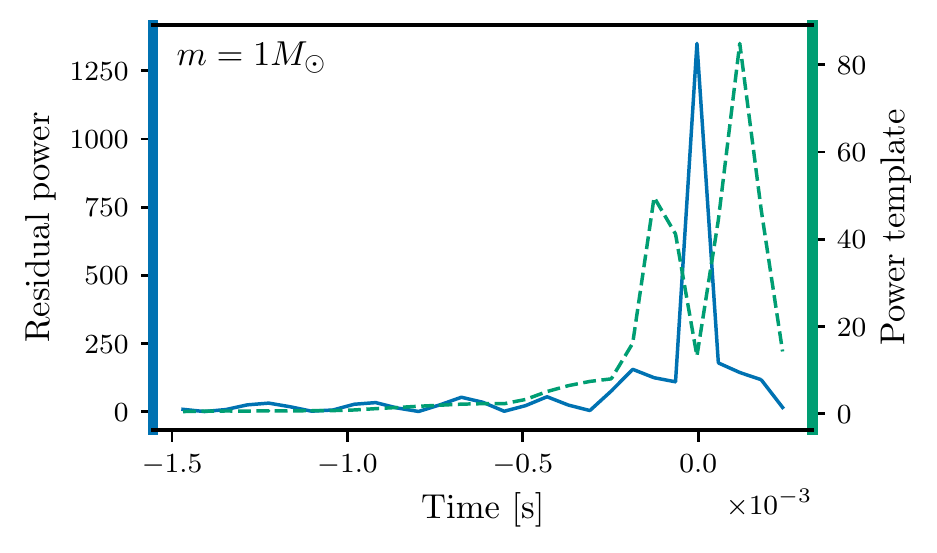}
	\end{minipage}
	\hfill
	\begin{minipage}{0.48\textwidth}
	\includegraphics{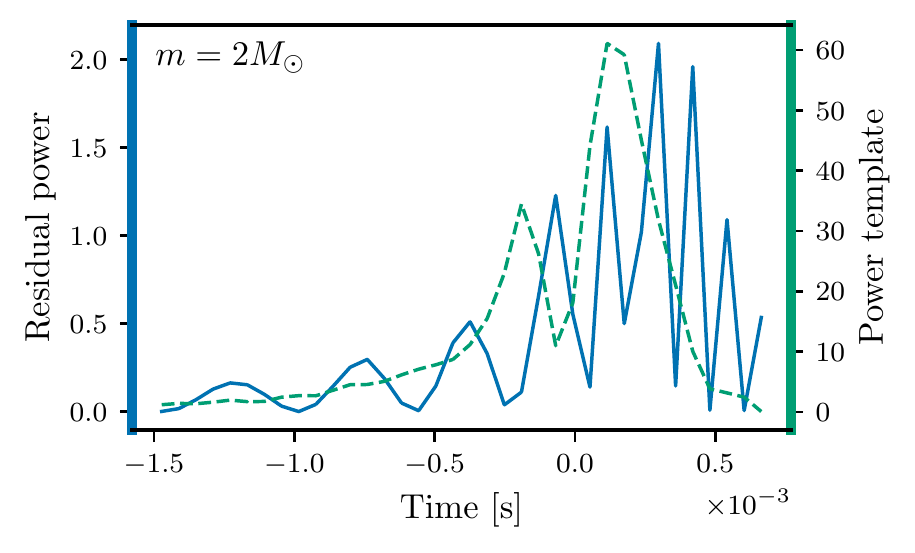}
	\end{minipage}
	\includegraphics{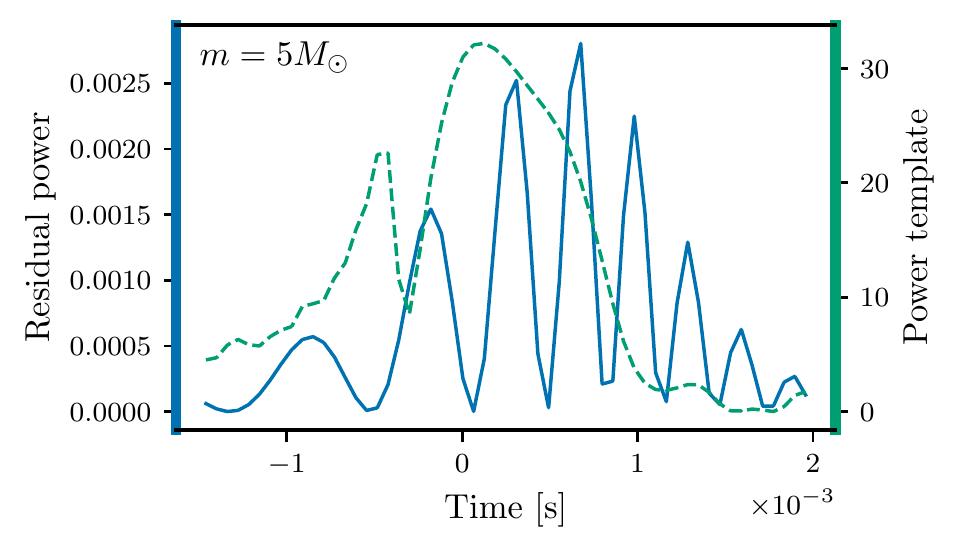}
	\caption{Residual power and template functions. The residual signals are  
		obtained from subtracting the GR waveforms from the EFT
		waveforms with the same source parameters.
		They are plotted
		with blue, solid lines (left axis). The template 
		functions $\tilde{Z}_{1}$ are computed from the GR signal using a
		value of $\tau$ equal to the sampling interval. They are
		plotted with dotted green lines (right axis). The
		template in Eq.~\eqref{eq:toy_model_expansion} is unmodeled, and is
		not expected to match actual signal at finite order, but it
		does capture the time interval over which the most power is
		present. }
	\label{fig:cc_basis}
\end{figure*}

We perform the same analysis as in Sec.~\ref{results_1} on the NR
waveforms motivated by the EFT. The strain signal $s(t)$ is now the full
numerical waveform obtained with the source at
$d_l=100$ Mpc. We assume that the standard waveform corresponding to the
best-fit parameters, $s_\text{SM}(t)$, is the NR waveform obtained with
the same source parameters in GR. These correspond to equal mass, non-spinning, quasi-circular black hole binaries with individual masses
given by $m/M_{\odot}= 1, 2 ~~\mbox{and}~~ 5.$ 

In Sec.~\ref{results_1}, the BM signature introduced was directly proportional to the 
template function, $Z_{1}$. As described in Sec.~\ref{formalism}, more realistic
residual signals such as the one obtained from the EFT-inspired waveform will require a
possibly infinite number of higher-order terms to be accurately described. The first
order (normalized) template function $\tilde{Z}_{1}$ obtained from the GR
waveforms using $\tau$ equal to the sampling interval is plotted for different
$m$ with dotted lines in Fig.~\ref{fig:cc_basis}. The power of the residual
strains is plotted with full blue lines. Although the template functions do not
track the short-scale time evolution, they do capture the time interval over
which residual power is present. 

To choose the timescale over which to cross-correlate, we note that the power
templates have two peaks. Ideally, one would integrate over the smaller feature
seen in the template---in this case, the narrower of the peaks in power. This
would be the feature seen around $-0.2$ to 0 milliseconds for $m=1 M_{\odot}$,
$-0.4$ to 0 milliseconds for $m=2 M_{\odot}$, and from $-0.6$ to $0.2$
milliseconds for  $M=5M_{\odot}$. However, for $m=1 M_{\odot}$ and  $2 M_{\odot}$, at the
maximum currently available sampling rate ($16384$Hz), the residual power
signals only span 2-5 sample points. We therefore choose to lose precision in
favor of a higher SNR: averaging over both peaks (up to the value of $\tau$
at which the mean power per bin $P$ scales as $\tau^{-1/2}$), which gives
the highest SNR. This corresponds to integrating from around $-0.25$  to $0.25$
milliseconds and from $-0.5$ to 0.5 milliseconds for $m=1 M_{\odot}$ and  $2 M_{\odot}$,
respectively. For $m=5 M_{\odot}$, the larger peak is smoother. As a result,
integrating over both peaks (for example, from $-1$ to 1 milliseconds) does not
give a larger SNR. For all masses, lower values of $\tau$ lead to a decrease in
SNR.

The values of $\tau$ chosen for each mass are shown in Table~\ref{tab:nr_results}. The coefficients obtained
by projecting the residual power onto the template with the aforementioned
values of $\tau$ are also shown in the table.

\subsection{Recovery of injected signal}

\begin{figure}[h]
	\includegraphics{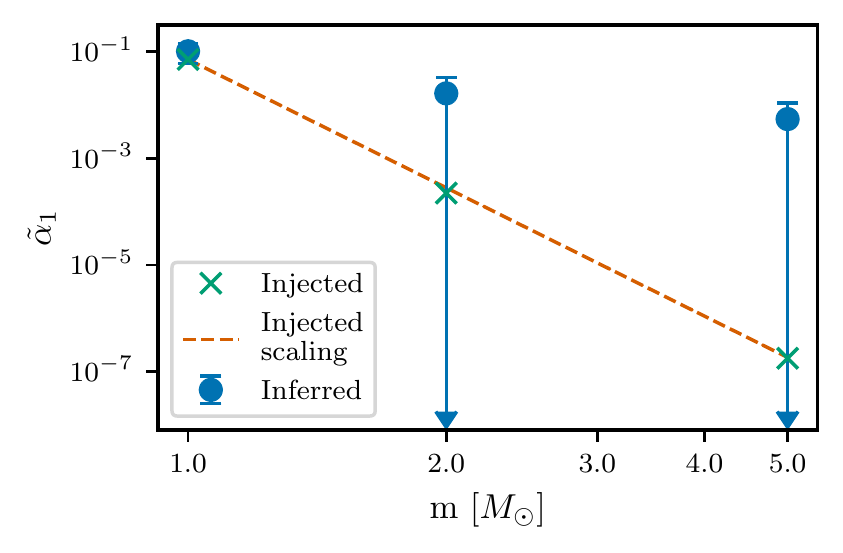}
	\caption{Injected and recovered values of $\tilde{\alpha}_{1}$. The
	projections of the actual residual signal on $Z_{1}$ are shown with
	green crosses. They correctly scale as $m^{-8}$, shown by the orange dotted line.
	The blue discs are values measured from toy data constructed with
	$500$ events. The error $\tilde{K}_{\tilde{\alpha}}$, shown by the
	error bars is several orders of magnitude larger than the injected
	values for $m=2$ and $5$ M$_\odot$. The measurements for these masses are consistent
	with zero, as indicated by the downward porting arrows, and the scaling
	with $m$ cannot be recovered. }
	\label{fig:alpha_scaling}
\end{figure}

The values of $\tilde{\alpha}_{1}$ corresponding to the injected residual power are
plotted with green crosses in Fig.~\ref{fig:alpha_scaling}. The values recovered
from toy data with $N_\text{GW} =500$ are plotted with blue discs and 
error bars. For $m=2$ and $5$ M$_\odot$, the recovered values are consistent with zero {in comparison to the 
standard deviation} $\tilde{K}_{\tilde{\alpha}_1}$,  as indicated by a downward arrow
on the error bars.

\begin{table*}[t]
	\begin{tabular}{ccccccc}
		\\[-1.5ex] \hline \hline \\[-1.5ex]
		$\frac{m}{M_{\odot}}$ & $\tau$ (s) &
		$\tilde{\alpha}_{1,\text{inj}}$  & $\tilde{\alpha}_{1,MLE}$ &
		SNR & $\mathcal{B}^\text{BM}_\text{SM}$ & FAR
		\\
		\hline
		1 & $4.27 \times 10^{-4}$  & $7.20\times 10^{-2}$ & $1.02\times
		10^{-1}
		$ & 
		{3.05}
		&
		{8.69}
		&
		{$1.6 \times 10^{-3}$}
		\\
		2 & $1.22 \times 10^{-3}$  & $2.22 \times 10^{-4}$ & $1.66
		\times 10^{-2}$ &
		{0.60}
		&
		{0.06}
		&
		{$2.7 \times 10^{-1}$}
		\\
		5 & $1.22 \times 10^{-3}$  & $1.78 \times 10^{-7}$ & $5.47
		\times 10^{-3}$ 
		&
		{0.46}
		&
		{0.02}
		&
		{$3.2 \times 10^{-1}$}
		\\
		 \hline \hline \\[-1.5ex]
	\end{tabular}	
	\caption{
	{
		Projection coefficients of the injected {BM} signal and
	the recovered {maximum likelihood estimation (MLE)}  of the NR EFT waveforms for $500$ events. The
	values of $\tau$ are chosen so that the SNR is maximized and the smallest scale feature captured. They depend on
	the best-fit waveform. 
}
The drop in Bayes factor from $m=1M_{\odot}$ to $m=2,5M_{\odot}$ is due
to the mean recovered value of $\tilde{\alpha}_{1}$ moving within error
range of $\tilde{\alpha}_{1}=0$. The False Alarm Rate (FAR) is computed
as the probability of obtaining a value equal or larger than the
$\tilde{\alpha}_{1}$ measured when there is no signal in the data (see
Appendix~\ref{app:far}) for a Gaussian stationary noise.
} \label{tab:nr_results}
\end{table*}

In this particular BM model, the power of the residual signal decreases with
mass as $m^{-8}$. Regardless of the template chosen, the projection of the
residual power onto it will also scale as $m^{-8}$. This is the case
for the injected values of $\tilde{\alpha}_{1}$. 
{
The dotted orange line
corresponds to a fit though these points with gradient $m^{-8}$---the template
functions correctly capture the scaling of the NR waveform residuals. The fact
that the template functions correctly capture the timescale and the power
scaling of the BM signature is not trivial, as they are agnostic to the model
and were not informed about the nature of the EFT used to simulate the NR
waveforms.
} {This is one of the salient aspect of the method \texttt{SCoRe}.}
For $m=2$ and $5$ M$_\odot$,
 $\tilde{\alpha}_{1,\text{inj}} \ll \tilde{K}_{\tilde{\alpha}}$. The Bayes
 factor for both these masses is inconsequential, and we
 cannot recover the scaling with mass.

We have assumed that the only term scaling with the source
parameter we are investigating, which is mass in this example,  is due to the
BM signature, so that the residual strain scales with a power $k$, {where $k$
captures the dependence of the BM signal}, on the source parameters $\theta$ as
$\Delta d \propto \theta^{k}$. For the EFT model considered here, $k=-4$ and
$\theta$ is the source mass. In reality, the residual strain may contain
contributions from unaccounted BM physics, waveform systematics, or a stealth
bias, for example. These contributions may scale with the same source parameter
as the BM signature, but with a different dependence, such as $\theta^{k'}$.
Then depending on the value of $k'$ relative to $k$ and the range over which
the value of $\theta$ is observed, one may still fail to recover the scaling
with $k$ even with large network SNR and $\tilde{K}_{\tilde{\alpha}}\ll
\tilde{\alpha}_{1,\text{inj}}$. In the case of a stealth bias, we would
na\"ively assume $k'=-4$. This is because, as the magnitude of the BM signature
changes, so will the magnitude of its projection onto the SM manifold. More
complicated dependence on the source parameters of the BM signature may yet
arise, as noted in~\cite{okounkovaGravitationalWaveInference2022}.

\section{Conclusion and future outlook}\label{conc}
In this work we propose a new method, \texttt{SCoRe}, to search for BM signatures
in GW data by projecting the cross-correlated residual power
spectrum signal on a theory-independent or theory-dependent template, which can
capture a large class of BM scenarios and its dependence on the GW source
parameters. 

As one of the key features of BM signatures is that they will be universal in all the
detectors, cross-correlating between different pairs helps mitigate
contamination from uncorrelated noise artifacts in the BM signal.
Correlated and strong transient noise, as well as any unmodeled BM physics,
will not be reduced by cross-correlation. To account for this, the second step
of the \texttt{SCoRe} method is to project the cross-correlated residual power
onto a template capturing a specific frequency dependence. This frequency
dependence can be tuned for both modeled and unmodeled searches. An expected
facet of BM signatures, however, is that they can be understood as a change in
the orbital energy radiated by the coalescing binary. As an alternative to
modeled searches, we therefore also propose to search for dependence on the
derivatives of the logarithm of orbital frequency as a tracker of energy radiated
per orbit.

Many BM signatures are expected to vary as a function of the source
parameters. The final step of the \texttt{SCoRe} method is to check for
correlation between the projection coefficients measured and the values of the
best-fit parameters used to compute the residuals. We incorporated a
hierarchical Bayesian framework for the marginalization of source parameters
(around a best-fit value) of multiple events of the cross-correlation signal
obtained by combining all possible pairs of GW detectors. 

For the unmodeled
searches, the template may not be able to completely capture the deviation from
the standard model, however, it can indicate signatures of possible deviations
from the standard model. To understand which BM scenarios best explain
the data, one can do a Bayesian model comparison. This can be incorporated into
the framework proposed in this work. Following that a model-dependent
search using the template choice which is tuned for the BM scenario which shows
maximum Bayesian evidence in the unmodeled search can be performed.    

To illustrate our method,
we have applied it to a toy model and also to waveforms
motivated by an EFT of
gravity~\cite{DeRham:2018bgz,Cardoso:2018ptl,Cano:2021myl,Cayuso:2020lca}. In
both cases, the injected residual power was {mainly} concentrated in the last
{few milliseconds} (before the peak of the waveform) of a signal and the noise
was Gaussian, stationary, and colored according to the expected O4 PSD.

For the toy model, we showed how well different signal strengths could be recovered, as measured by
the Bayes factor in favor of the presence of a BM signature, for different
levels of power injected and number of events combined. The events used all had
the same source parameters. Crucially, they were all at a luminosity distance $d_l$
of $100$ Mpc. In reality, sources are expected to be distributed over a large
range of luminosity distances, in which case the residual power will scale as
$d_l^{-2}$. In addition, since the properties of the source and their
astrophysical population will critically impact the Bayes factors recovered,
we will work next on taking into account astrophysical populations to obtain
more realistic estimates on how well the method can infer the presence and
structure of BM signatures.

The illustration of the method on the EFT-motivated NR waveforms constrained the
BM signatures around zero and could not detect the injected signatures for the
masses giving the smallest residual power ($m=2,5M_{\odot}$). However, the
model-independent template we suggest for agnostic searches was successful in
capturing the time range over which most of the residual power was emitted. It
could also correctly infer the minute BM signal for
$m=M_{\odot}$.
As this search is performed using a model-independent template, it
could not reproduce the exact structure of the residual signal, as expected. In
this scenario, we suggest that, once one has identified a promising result
using the model-independent search, the templates should be refined. This can
either be done by using a model-dependent templates or by adding higher-order
derivatives to the model-independent template functions.

This work has been to present the main lines of the \texttt{SCoRe} method. One
of the main caveat of our results is that they assume Gaussian, stationary
noise. Although these assumptions are used in the LVK
analyses~\cite{theligoscientificcollaborationGWTC1GravitationalWaveTransient2019,abbottGWTC2CompactBinary2021,theligoscientificcollaborationGWTC3CompactBinary2021},
we are working on applying this method on the GW data available to
investigate the effect of realistic noise, and what the FAR for different
templates may be. We will also consider, in future work, how transient noise
(glitches) affects the recovery of BM signals from data.

We hope that \texttt{SCoRe} will not only be useful to search for signatures of
new physics, but also a large class of scenarios within the standard model
ranging from unmodeled GR effects to deviation in the waveform due to waveform
modeling systematics. 

We have illustrated the method with toy models in this work and will,
in future works, provide tests and predictions for its applications.
We will explore the capability of this
technique in recovering different scenarios and the corresponding appropriate
template banks which can be useful to search for both modeled and unmodeled
searches. 

\section*{Acknowledgements}
The authors are thankful to Nathan K. Johnson-McDaniel for reviewing the manuscript as a part of the LVK publication and presentation procedure and giving useful comments. We are also grateful to Reed Essick for useful discussions regarding residual
tests. We also acknowledge Vitor Cardoso, Katherina Chatzioannou, Will Farr,
Max Isi, Sean McWilliams, Rafael Porto and the LIGO-Virgo-KAGRA Scientific Collaboration TGR group for discussions during the course of this work.
The authors would like to thank the LIGO-Virgo-KAGRA Scientific Collaboration for providing the noise curves. 
This research has made use of data or software obtained from the Gravitational Wave Open Science Center (gw-openscience.org), a service of LIGO Laboratory, the LIGO Scientific Collaboration, the Virgo Collaboration, and KAGRA. LIGO Laboratory and Advanced LIGO are funded by the United States National Science Foundation (NSF) as well as the Science and Technology Facilities Council (STFC) of the United Kingdom, the Max-Planck-Society (MPS), and the State of Niedersachsen/Germany for support of the construction of Advanced LIGO and construction and operation of the GEO600 detector. Additional support for Advanced LIGO was provided by the Australian Research Council. Virgo is funded, through the European Gravitational Observatory (EGO), by the French Centre National de Recherche Scientifique (CNRS), the Italian Istituto Nazionale di Fisica Nucleare (INFN) and the Dutch Nikhef, with contributions by institutions from Belgium, Germany, Greece, Hungary, Ireland, Japan, Monaco, Poland, Portugal, Spain. The construction and operation of KAGRA are funded by Ministry of Education, Culture, Sports, Science and Technology (MEXT), and Japan Society for the Promotion of Science (JSPS), National Research Foundation (NRF) and Ministry of Science and ICT (MSIT) in Korea, Academia Sinica (AS) and the Ministry of Science and Technology (MoST) in Taiwan. This material is based upon work supported by NSF's LIGO Laboratory which is a major facility fully funded by the National Science
Foundation. 
This research was supported in part by a NSERC Discovery Grant and CIFAR (LL),
the Simons Foundation through a Simons Bridge for Postdoctoral Fellowships  (SM) as well as Perimeter Institute for Theoretical Physics. 
Research at Perimeter Institute is supported by the Government of
Canada through the Department of Innovation, Science and Economic Development
Canada and by the Province of Ontario through the Ministry of Research,
Innovation and Science. The work of SM is a part of the $\langle \texttt{data|theory}\rangle$ \texttt{Universe-Lab} which is supported by the TIFR and the Department of Atomic Energy, Government of India.  In the analysis done for this paper, we have used the
following packages: \textsc{PyCBC}~\cite{nitzGwastroPycbcV22022}, \textsc{LALSuite}~\cite{lalsuite}, 
\textsc{NumPy}~\cite{harris2020array}, \textsc{SciPy}~\cite{2020SciPy-NMeth}
and \textsc{Matplotlib}~\cite{Hunter:2007} with
\textsc{Seaborn}~\cite{Waskom2021}.

\begin{appendices}

\section{Modeling residual strain}\label{app:strain}
In Sec. \ref{sec:formalism_basis}, we have discussed the template constructed to recover signal from the cross-correlation power spectrum. In this section, we will briefly describe the template that can be used to model any deviation on the strain.  
Instead of modeling the residual power, $\mathcal{S}_\theta(t)$, we can 
model the residual strain $\Delta s_\theta$. In that case, the above equations
written for $\mathcal{P}_\theta(t)$ will only get a minor modification.  The
decomposition of the signal in terms of template functions
(Eq.~\eqref{eq:cc_expansion}) becomes
\begin{equation}\label{eq:strain_expansion}
    \Delta s_{\theta_\text{MLE} }(t) 
    = 
    \sum_{i=1}^{i=n}
    \beta_{i}(\theta_\text{MLE})
    \zeta_{i}(f(t)),
\end{equation}
where the coefficients $\beta{i}$ and bases $\zeta_{i}$ give $\alpha_{i}$ and
$Z_{i}$ when cross-correlated using Eq.~\eqref{eq:cc_definition}. In this
case, both sides of the equation still depend on the timescale $\tau$, but they
take values in strain rather than strain power.

In analogy to Eq.~\eqref{eqprjc1}, we can
write the estimator for $\beta$ as  
\begin{equation}\label{eqprjc2}
    \hat \beta_i (\theta_\text{MLE}) 
    =
    \int_{t_s}^{t_e} dt
    W_{i}(t)
    D(t)^{1/2}
    \zeta_{i},
\end{equation}
with $W_{i}(t)= K^{-1}_n(t)/ \int_{t_s}^{t_e} K^{-1}_n(t)
\zeta_{i}(t)^2 dt$. The corresponding SNR is \begin{align}\label{totsnr2}
   \rho_{i}^2= \sum^{N_{\rm det}}_{x=1, x'>x} \sum^{N_{GW}}_j 
   \left(
   	\frac{(\hat \beta^{xx'}_i (\theta_\text{MLE}))^2}{{(K^{xx'}_{\beta_{i}})}^2} 
    \right) _j,
\end{align}
where, ${(K^{xx'}_{\beta_{i}})}^2$ is the noise on the parameter $\beta$ is defined as
\begin{equation}
   {(K^{xx'}_{\beta_{i}})}^2
   = 
	   \int_{t_s}^{t_e} dt 
   \left(
	   W_{i}(t)
	   K_{n}^{\frac{1}{2}}
	   \zeta_{i}(t)
    \right)^2,
	\label{noisebeta}
\end{equation}
where there is now a square root on $K_{n}$.

Again for a Gaussian, stationary, uncorrelated noise, we will get a
$\sqrt{N_{GW}}$ enhancement in the SNR by combining $N_\text{GW}$ events and number of detector pairs $\sqrt{N_{\rm det}(N_{\rm det} -1)/2}$.

\section{False Alarm Rate}\label{app:far}

\begin{figure}
	\includegraphics{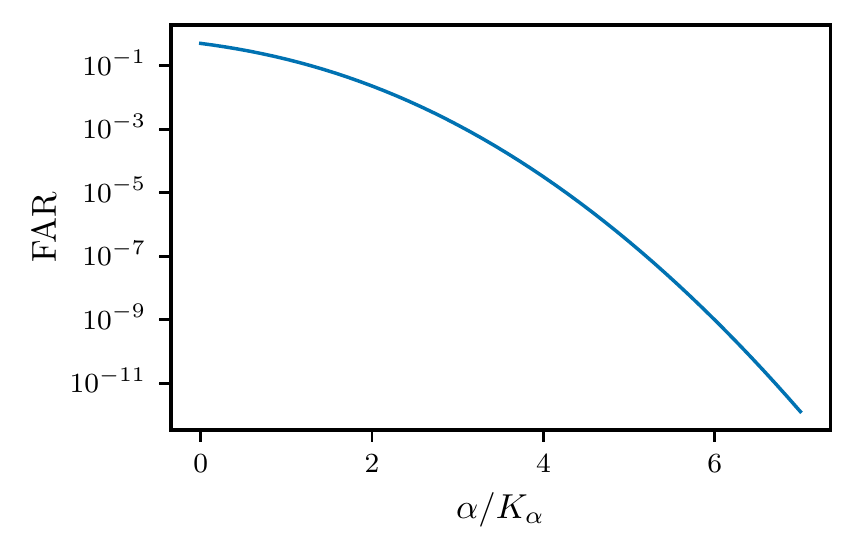}
	\caption{False alarm rate (FAR) on the measurement of $\alpha$ for Gaussian,
		stationary noise. We define the FAR for a value of the
		projection coefficient $\alpha$ as the probability of
		obtaining a measurement equal or greater to $\alpha$ from data
		without a signal. Since our noise is Gaussian and stationary,
		this is given by the survival function of a Gaussian
		distribution. Here, we plot the survival function for the
		normal distribution, which gives the FAR as a function of the
		ratio of $\alpha$ to the error on $\alpha$. 
}
		\label{fig:far}
\end{figure}
\begin{figure}
	\begin{minipage}{0.49\textwidth}
		\includegraphics{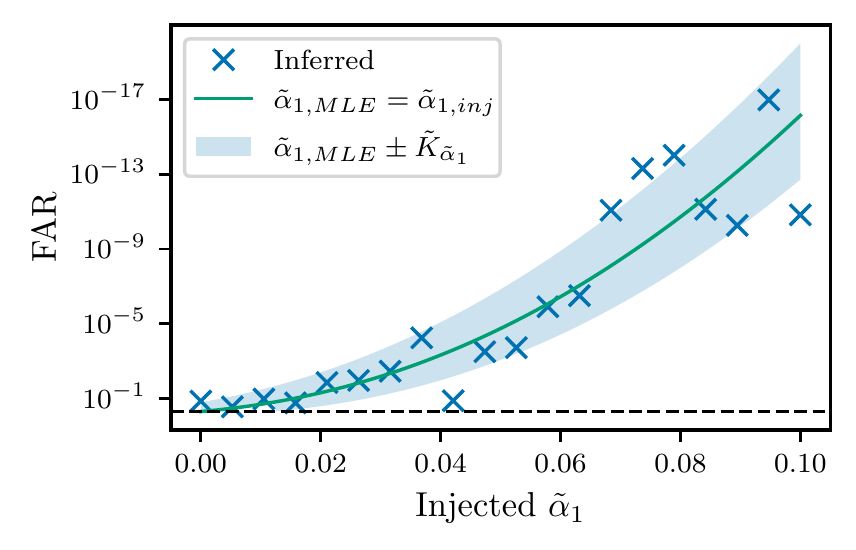}
	\end{minipage}
	\hfill
	\begin{minipage}{0.49\textwidth}
		\includegraphics{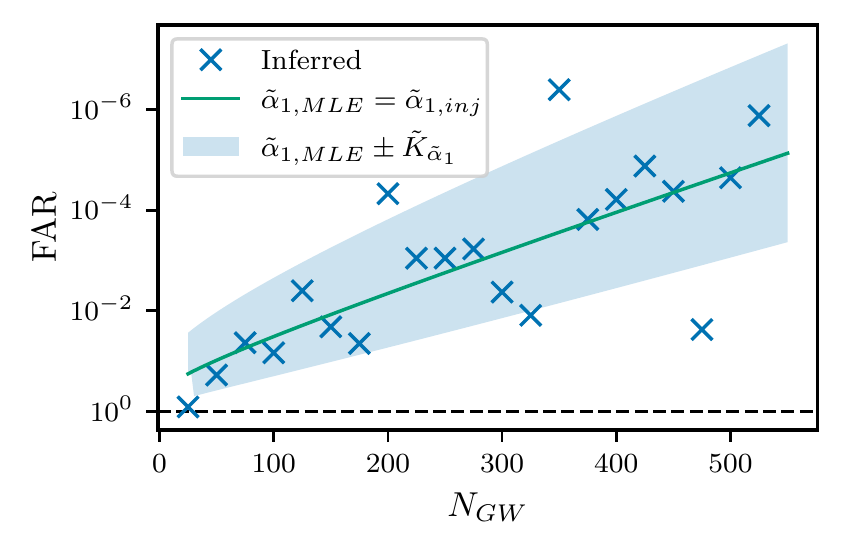}
	\end{minipage}
	\caption{False Alarm Rate (FAR), or the probability of measuring a
		projection coefficient equal or greater to $\tilde{\alpha}_{1}$
		. The y-axis scale has been inverted to show that the FAR
		scales in the same way as $\mathcal{B}_\text{SM}^\text{BM}$,
		the Bayes factor in favor of a BM hypothesis, shown in
		Fig.~\ref{fig:toy_model_bayes}. The dashed line is a visual aid
		indicating a FAR of 0.5.
	}
	\label{fig:far_scaling}
\end{figure}

The False Alarm Rate (FAR) is a measure of the probability of detecting a
signal when it is absent from the data. In the LVK
catalogs~\cite{theligoscientificcollaborationGWTC1GravitationalWaveTransient2019,abbottGWTC2CompactBinary2021,theligoscientificcollaborationGWTC3CompactBinary2021},
the FARs reported are computed using real data and using different methods
depending on the pipeline used. In this work, we
define the FAR for a given value of $\alpha$ to be the probability of obtaining
a value of $\alpha$ equal or greater when there is no signal in the data. As we are
considering Gaussian, stationary noise, the (frequentist) distribution of
$\alpha$ measurements from noise is a Gaussian distribution centred around
zero, with variance given by Eq.~\eqref{eq:sigma_2_alpha}. The FAR is then the
survival function of this Gaussian evaluated at $\alpha$: 
$$
\begin{aligned}
	\text{FAR}(\alpha)
	  &=
	  1
	  -
	  \int^{\alpha}_{-\infty}
	  d \alpha
	  \frac{1}{K_{\alpha} \sqrt{2\pi}}
	  \exp 
	  \left(
		  -
		  \frac{1}{2}
		  \left(
			  \frac{\alpha}{K_{\alpha}} 
		  \right)^{2}
	  \right) 
\end{aligned}
$$
The FAR for any pair of measured $\alpha$ and  $K_{\alpha}$ can be obtained by
substituting $\alpha/K_{\alpha}$ in the survival function for the
normal distribution, plotted in Fig.~\ref{fig:far}. The same can be done with 
$\tilde{\alpha}_{1}/\tilde{K}_{\alpha}$.

The FAR scales with the injected projected coefficient and the number of events
in the same way as the Bayes factor, as shown in Fig.~\ref{fig:far_scaling}.
In future work, we will compute the FAR by considering available GW
data and how often an event that is not believed to contain any BM
signal can lead to a non-zero measurement of the projection coefficients.

\clearpage

\end{appendices}

\def\urlprefix{}
\def\url#1{}
\bibliography{main.bib}

\end{document}